\documentclass{article}

\PassOptionsToPackage{numbers, square, compress}{natbib}

\usepackage[final]{neurips_2023}

\usepackage[utf8]{inputenc} % allow utf-8 input
\usepackage[T1]{fontenc}    % use 8-bit T1 fonts
\usepackage{hyperref}       % hyperlinks
\usepackage{url}            % simple URL typesetting
\usepackage{booktabs}       % professional-quality tables
\usepackage{amsfonts}       % blackboard math symbols
\usepackage{nicefrac}       % compact symbols for 1/2, etc.
\usepackage{microtype}      % microtypography
\usepackage{xcolor}         % colors
\usepackage{graphicx}
\usepackage{amsmath}
\usepackage[ruled,linesnumbered]{algorithm2e}
\usepackage{adforn}
\usepackage{multirow}
\usepackage{bm}

\title{Active Learning with Task Adaptation Pre-training for Speech Emotion Recognition}

\author{%
  Dongyuan Li \\
  Tokyo Institute of Technology\\
  \texttt{lidy94805@gmail.com} \\
\And
Ying Zhang \\
RIKEN \& Tohoku University \\
\texttt{zhang@lr.pi.titech.ac.jp} \\
\AND
Yusong Wang \\
Tokyo Institute of Technology \\
\texttt{wangyi@lr.pi.titech.ac.jp} \\
\And
Funakoshi Kataro \\
Tokyo Institute of Technology \\
\texttt{funakoshi@lr.pi.titech.ac.jp} \\
\And
Manabu Okumura \\
Tokyo Institute of Technology \\
\texttt{oku@lr.pi.titech.ac.jp} \\
}

\begin{document}

\maketitle

\begin{abstract}
Speech emotion recognition (SER) has garnered increasing attention due to its wide range of applications in various fields, including human-machine interaction, virtual assistants, and mental health assistance.
However, existing SER methods often overlook the information gap between the pre-training speech recognition task and the downstream SER task, resulting in sub-optimal performance. 
Moreover, current methods require much time for fine-tuning on each specific speech dataset, such as IEMOCAP, 
which limits their effectiveness in real-world scenarios with large-scale noisy data.
To address these issues, we propose an active learning (AL)-based fine-tuning framework for SER, called \textsc{After}, that leverages task adaptation pre-training (TAPT) and AL methods to enhance performance and efficiency.
Specifically, we first use TAPT to minimize the information gap between the pre-training speech recognition task and the downstream speech emotion recognition task.
Then, AL methods are employed to iteratively select a subset of the most informative and diverse samples for fine-tuning, thereby reducing time consumption.
Experiments demonstrate that our proposed method \textsc{After}, using only 20\% of samples, improves accuracy by 8.45\% and reduces time consumption by 79\%. 
The additional extension of \textsc{After} and ablation studies further confirm its effectiveness and applicability to various real-world scenarios.
Our source code is available on Github for reproducibility. (https://github.com/Clearloveyuan/AFTER).
\end{abstract}

\section{Introduction}\label{sec:Introduction}

\textit{The language of tones is the oldest and most universal of all our means of communication}~\cite{blanton}. Speech emotion recognition (SER) aims to identify emotional states conveyed in vocal expressions, making it an essential topic in tone and language analysis.~\footnote{This paper is an extended version of our preliminary paper~\cite{10389652} presented in the IEEE AUTOMATIC SPEECH RECOGNITION and UNDERSTANDING (IEEE ASRU 2023). Please refer to the detailed differences outlined in Section~\ref{sec:previous}}
It has garnered significant attraction in both the industrial and academic communities, including speech-to-text translation~\cite{DBLP:conf/iclr/TaghaviSS23,DBLP:journals/access/SantosoYIHM22}, dialogue system~\cite{bertero-etal-2016-real,10446571,li-etal-2023-joyful}, 
medical surveillance systems~\cite{clavel2008fear}, psychological treatments~\cite{elsayed2022speech,DBLP:conf/coling/LiYFO22}, and intelligent virtual voice assistants~\cite{la2020human,wang2023emp}.

With the development of deep learning techniques in natural language processing~\cite{zhang-etal-2021-language,YZ2024,wu2024new} and computer vision~\cite{wang2024insectmamba,li2024ecnet,yang2024comparative}, many SER methods have been proposed.  These SER methods are broadly classified into classic machine learning-based methods and deep learning-based methods~\cite{abbaschian2021deep}. 
The former methods~\cite{GharsellaouiSY19,skevopoulos19} typically consist of three main components: feature extraction, feature selection, and emotion recognition. However, selecting and designing features for specific corpora is time-consuming~\cite{AyadiKK11}, and they consistently exhibit poor generalization on unseen datasets~\cite{PadiSSM21}. Deep learning-based methods can address these issues by automatically extracting more abstract features to improve generalization~\cite{LeCunBH15,DBLP:conf/icassp/MoraisHZGDA22,DBLP:journals/corr/abs-2307-10757}. They benefit from various neural network architectures such as convolutional neural networks (CNNs)~\cite{Aftab} and transformers~\cite{GhrissYRSW22}.
With the development of pre-trained language models~\cite{kenton2019bert} and the availability of large-scale datasets, various pre-trained automatic speech recognition (ASR) models have been proposed. 
ASR models, in this draft, refer to those models that use machine learning or artificial intelligence technology to process human speech into readable text such as wav2vec 2.0~\cite{baevski2020wav2vec}, HuBERT~\cite{babu2021xls} and Data2vec~\cite{baevski2022data2vec}.
These ASR models use speech's acoustic and linguistic properties to provide more robust and context-aware representations for speech signals. 
Xia et al.\citep{XiaCRS21} proved that fine-tuning SER datasets on wav2vec 2.0~\cite{SchneiderBCA19} obtain state-of-the-art (SOTA) performance on IEMOCAP~\cite{BussoBLKMKCLN08}. This finding has inspired researchers to explore new fine-tuning strategies on ASR models, becoming a new paradigm for SER.
For example, \cite{abs-2210-14636} proposed a self-distillation SER model to fine-tune wav2vec 2.0 obtaining SOTA performance on the DEMoS dataset~\cite{Cabaleiro20}. And \cite{2207-14418} fine-tuned wav2vec 2.0 by jointly optimizing the SER and ASR tasks achieving SOTA performance in Portuguese datasets.

Although the aforementioned methods achieve considerable success, several issues still need to be addressed. \textbf{(1)} Current methods seldom consider the information gap between the pre-trained ASR and downstream SER tasks. For example, wav2vec 2.0~\cite{baevski2020wav2vec} adopts a masked learning objective to predict missing frames from the remaining context, while the downstream SER~\cite{Aftab,BaruahB22} task aims to minimize cross-entropy loss between predicted and referenced emotion labels for speech signals.
Suchin et al.~\cite{gururangan-etal-2020-dont} proved that the information gap would decrease the performance of downstream tasks.
To address it, Pseudo-TAPT~\cite{2110-06309} first uses K-means to obtain pseudo-labels of speech signals and uses supervised TAPT~\cite{gururangan-etal-2020-dont} for continual pre-training.
However, K-means is sensitive to the initial value, making Pseudo-TAPT unstable and computationally expensive.
\textbf{(2)} Current methods only fine-tune and validate the performance on a specific speech dataset. For example, \cite{XiaCRS21} train their models solely on the IEMOCAP, leading to over-fitting and poor generalization for unseen datasets. 
Real-world scenarios contain much heterogeneous and noisy data, which hinders the application of these SER methods.
Heterogeneous means that real-world scenarios should contain different voice background, languages, devices for recording speech, and speech types (spontaneous speech and acted speech).
Please note that ``noisy data'' does not refer to acoustically noisy data (unclear speech or unrecognized audio). We define ``noisy data'' as the specific noise in the speech emotion recognition task, including outliers and redundant samples. Specifically, outliers encompass various ambiguous emotions due to the complexity of speech, which can lead to inaccurate emotional annotations and degrade the performance of the model. Redundant samples being trained repeatedly does not improve the model's accuracy. Instead, they lead to an uneven distribution of data, making it more challenging to identify emotions with a limited amount of data.
\textbf{(3)} Pre-trained ASR models often contain millions of parameters, for example, wav2vec 2.0 contains 317 million parameters, which is time-consumption for real-world and large-scale datasets.

To address the aforementioned issues, we propose an active learning-based fine-tuning framework for SER, referred to as \textsc{After}, which can be easily applied to noisy and heterogeneous real-world scenarios.
Specifically, we first propose an unsupervised task adaptation pre-training (TAPT) method~\cite{gururangan-etal-2020-dont} to reduce the information gap between the pre-trained and downstream SER tasks, enabling the pre-trained model to understand the semantic information of the SER task. 
Then, we create two large-scale heterogeneous and noisy datasets to simulate real-world scenes. Furthermore, we propose AL strategies with clustering-based initialization to iteratively select a smaller, more informative, and diverse subset of samples for fine-tuning. This approach can efficiently eliminate noise and outliers, improve generalization, and reduce time consumption.

Our main contributions can be summarized as follows:
\begin{itemize}
    \item To the best of knowledge, we are the first to propose a general task adaptation pre-training and active learning-based fine-tuning framework for the speech emotion recognition task to address the information gap, noisy sensitive, and low efficiency issues.
    \item We created three additional large-scale speech emotion recognition datasets to simulate different complex real-world scenarios by merging existing high-quality speech emotion datasets. These datasets represent noisy and heterogeneous real-world situations. And we released our created datasets on Github link to share with other researchers.
    \item Extensive experiments demonstrate the effectiveness and efficiency of our proposed methods \textsc{After}, and we perform well on IEMOCAP, Merged Dataset, and Merged-2 Dataset with four emotional categories, as well as SAVEE and Merged-3 Dataset with seven emotional categories. Additional extensions of \textsc{After} and demonstrate the effectiveness and applicability. 
\end{itemize}

The remainder of this paper is organized as follows. 
In Section~\ref{sec:Related_Work}, we provide a literature review of the most related work, including speech emotion recognition, active learning, and task-adaptation pre-training. 
In Section \ref{sec:Method}, we carefully introduce our methodology: \textsc{After}, in detail. 
In Section~\ref{sec:experiment1}, we describe the experimental corpora and setup in detail. 
In Section~\ref{sec:experiment2}, we present our experimental results and analyses. 
We present the limitations of this study in Section~\ref{sec:limitation} and the differences with the previous conference version in Section~\ref{sec:previous}.
Finally, we give the conclude and discuss future work of this study in Section~\ref{sec:conclusion}.

\section{Related Work}\label{sec:Related_Work}

\subsection{Speech Emotion Recognition (SER)}

The SER task is one of the key components in human-machine interaction and human
communication systems~\cite{DBLP:journals/taffco/LatifRKJQS23}.
With the development of deep learning~\cite{lai2024language,lai2024adaptive,Ying_Zhang2023}, several attempts have been made to automatically learn emotion representations from audio signals using neural networks~\cite{DBLP:conf/icassp/ChangRNQS23,DBLP:conf/icassp/ChenXXPD23,DBLP:conf/icassp/DangVNW23}.
However, commonly used SER datasets, such as MSP-Podcast~\cite{DBLP:conf/icassp/SrinivasanHK22}, IEMOCAP~\cite{BussoBLKMKCLN08} and CMU-MOSEI~\cite{DBLP:conf/acl/MorencyCPLZ18}, are relatively small compared to automatic speech recognition datasets. This limitation restricts the ability for pre-trained ASR models to improve the accuracy of emotion recognition. Self-supervised pre-trained models, such as Transformers, provide a solution by first learning from a large-scale speech corpus without explicit labeling~\cite{BaruahB22,DBLP:conf/interspeech/DissanayakeSSWN22}. The knowledge learned from pre-training datasets can be transferred to downstream tasks by either using the model as a feature extractor~\cite{Lavania2023} or directly fine-tuning the whole model~\cite{chen2023pre}. 
Please note that a model trained by a self-supervised learning algorithm is called a self-supervised learning (SSL) model in speech research.
While initially introduced for natural language processing (NLP), several SSL-based pre-trained ASR models have been developed for speech processing, including wav2vec 2.0~\cite{baevski2020wav2vec}, HuBERT~\cite{Omar}, and CLAP~\cite{DBLP:conf/icassp/WuCZHBD23}.
Taking wav2vec 2.0~\cite{baevski2020wav2vec} as an example, which serves as the base model in this draft, it comprises a multi-layer convolutional neural network (CNNs) designed to predict future frames based on past frames, achieving through the minimization of a contrastive loss. Additionally, wav2vec 2.0 utilizes a transformer-based architecture, employing a masked learning objective to predict missing frames within the given context.
These pre-trained models consistently demonstrate state-of-the-art performance across various SER datasets. 
For instance, \cite{DBLP:journals/corr/abs-2011-05585} observe that wav2vec 2.0 features surpass traditional spectral-based features in SER applications. \cite{2110-06309} showcase the advantages of task-adaptive pre-training in the wav2Vec 2.0 model, leading to a significant improvement in overall model performance.
Furthermore, \cite{XiaCRS21} conduct a comparative analysis of features extracted with different temporal spans, concluding that features with longer temporal context, such as those of wav2vec 2.0, exhibit superior performance in SER. 
\cite{DBLP:conf/interspeech/PepinoRF21} demonstrate that the features derived from a linear combination of layers outperform single-layer representations in wav2vec 2.0 for SER applications.

While these studies demonstrate the usefulness of pre-trained models as feature extractors, little research has been done on how to efficiently fine-tune them for SER. 
Different from the above-mentioned works, we focus on proposing a general fine-tuning framework to apply effectively and efficiently to any type of pre-trained ASR models for SER tasks.

\subsection{Active Learning}

Active learning is an extensively research challenge in the field of machine learning, encompassing a variety of scenarios and query strategies~\cite{DBLP:conf/icimcs/XuSZ13,zhang-etal-2022-survey}.
In recent years, there has been a resurgence of interest in active learning within the NLP community~\cite{zhang-etal-2022-survey}. Recent studies have employed active learning with BERT models for specific tasks such as intent classification~\cite{10.1145/3374587.3374611}, sentence matching~\cite{bai-etal-2020-pre}, parts-of-speech tagging~\cite{DBLP:journals/tacl/ChaudharySAN21}, and named entity recognition~\cite{DBLP:journals/npl/LiuTZSXW22}. 
\cite{MargatinaVBA21} advocate for continued pre-training on unlabeled data in the context of active learning. \cite{DBLP:journals/tacl/RotmanR22} adapt active learning for multi-task scenarios involving transformer models. \cite{ein-dor-etal-2020-active} conduct an extensive empirical study of existing active learning strategies on binary classification tasks. 
\cite{yuan-etal-2020-cold} adapt the BADGE~\cite{DBLP:conf/iclr/AshZK0A20} framework for active learning with BERT. While BADGE computes gradient embeddings from a neural network's output layer and subsequently clusters the gradient space.

To the best of our knowledge, our work is the first active learning-based fine-tuning framework in the speech domain. Instead of focusing on proposing complex active learning query strategies, we concentrate on evaluating the effectiveness of active learning for SER.
Through experimentation, in Section~\ref{sec.compare with sota}, we validate its effectiveness and aspire to propose more efficient methods to advance this task in the future.

\subsection{Task Adaptation Pre-training}

Task-adaptive pre-training (TAPT) is a significant area of research, as introduced by \cite{gururangan-etal-2020-dont}. Essentially, TAPT involves customizing a language model (LM) for a specific task, leading to improved model performance.
\cite{gururangan-etal-2020-dont} explore the benefits of tailoring a pre-trained model like RoBERTa to a specific task domain. They investigate four distinct domains, covering biomedical and computer science publications, news, and reviews, and spanning eight classification tasks. Their exploration extended to assessing the transferability of adapted language models across different tasks and domains. Additionally, they conduct a study to evaluate the importance of pre-training on human-curated data.
\cite{DBLP:conf/chr/KonleJ20} discuss various strategies for adapting BERT and DistilBERT to historical domains and tasks in computational humanities. The outcomes support the idea of continuous pre-training in machine learning tasks to enhance performance stability. A combined approach of domain adaptation and task adaptation shows positive effects. Task adaptation alone is versatile and applicable in various setups, unlike domain adaptation, which requires a substantial amount of in-domain data.

Several approaches have been explore to make TAPT more efficient, especially with methods involving word embeddings.
For example, \cite{DBLP:conf/acl/NishidaNY21} propose TAPTER, enhancing pre-trained language model embeddings for domain adaptation. It outperforms standard methods when in-domain data is limited.
\cite{elboukkouri:tel-03560502} advocate re-training from a general model for low-resource scenarios, yielding comparable performance with slight trade-offs.
\cite{DBLP:conf/emnlp/SachidanandaKL21} adapt tokenizers to transfer pre-trained models to new domains, achieving over 97\% performance benefits but introducing a 6\% increase in model parameters.

In this study, we introduce a straightforward approach for continuous training of a pre-trained model with a task-related loss function on downstream tasks. Our experimental results, detailed in section~\ref{ablation_study}, demonstrate the effectiveness of this method.

\section{Methodology}\label{sec:Method}

% \section{Methodology}

\begin{figure*}[h]
\includegraphics[width=1\textwidth]{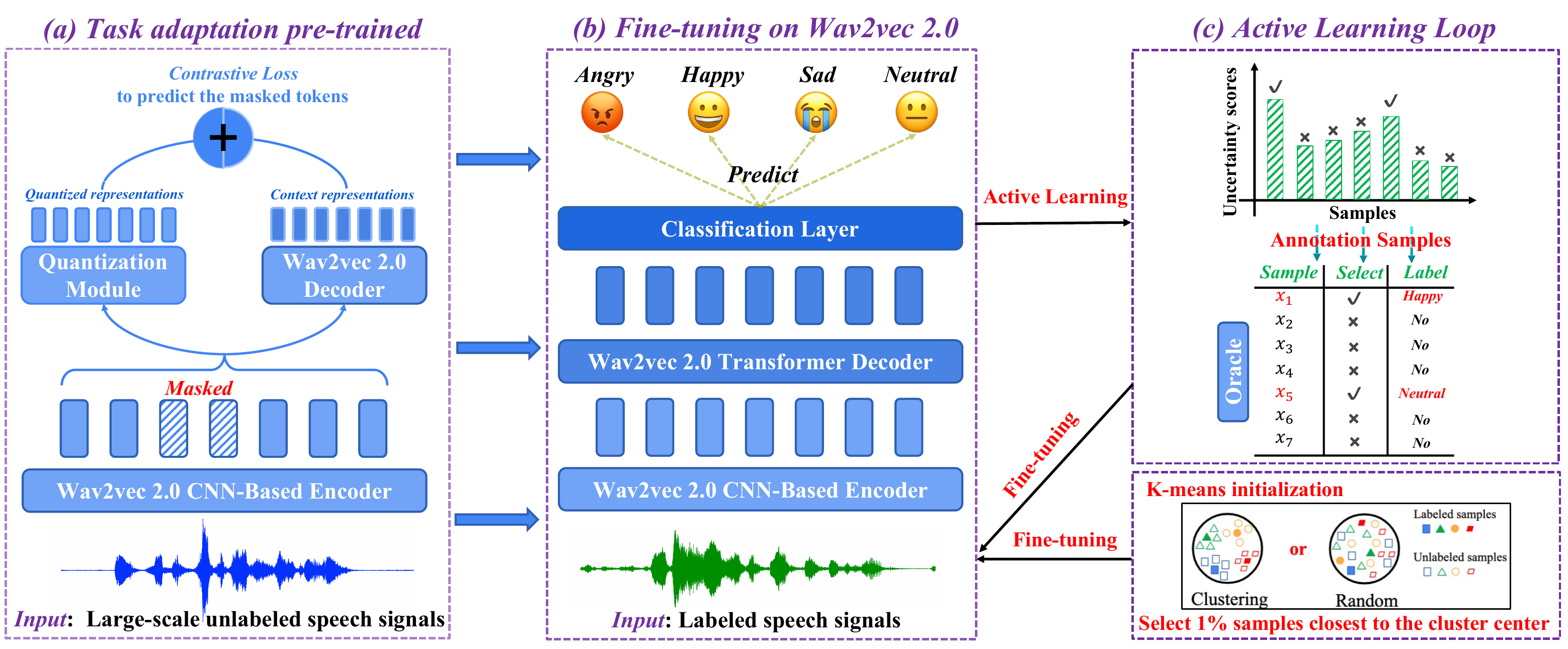}
\centering
\caption{Model overview. First, we pre-train an off-the-shelf wav2vec 2.0 in the TAPT manner. Then, we adopt an AL method to select unlabeled samples for iterative annotation. These labeled samples are used to fine-tune the wav2vec 2.0 model for SER.}
\label{overall}
\end{figure*}

The overall framework of \textsc{After} is depicted in Figure~\ref{overall}, comprising three main components: a \textit{task adaptation pre-trained} module, an \textit{active learning-based fine-tuning} module, and an \textit{emotion classification} module. 
First, we will formally define the task of SER, and subsequently introduce each component of \textsc{After} in detail.

\subsection{Notations and Task Formulation}
Given a speech dataset $\mathcal{D}$ = $\{(\mathbf{x}_{i},y_{i})\}_{i=1}^{N}$ where $\mathbf{x}_{i}$ represents the $i$-th speech signal and $y_{i}$ represents its corresponding emotion label, 
we aim to fine-tune a pre-trained automatic speech recognition model $\mathbf{M}$, such as wav2vec 2.0~\cite{2110-06309}, on the labeled speech datasets $\mathcal{D}_{\text{train}}$ to obtain accurately predicted emotion labels for all speech signals.

\subsection{Task Adaptation Pre-training (TAPT)}

As shown in Figure~\ref{overall} (a), we introduce the TAPT component in detail.
To better leverage pre-trained prior knowledge for the benefit of downstream tasks and minimize the information gap between pre-trained tasks and downstream tasks, \cite{gururangan-etal-2020-dont} continued training the pre-trained model RoBERTa~\cite{Roberta} on downstream datasets via the same loss of the pre-training task (reconstructing the masked tokens of input sentences, similar to BERT~\cite{kenton2019bert}), resulting in significant improvements in downstream text classification tasks.
Inspired by their work, we added an additional step into \textsc{After} by continuing training the pre-trained automatic speech recognition model, using wav2vec 2.0 as an example in this study, on downstream training datasets for the speech emotion recognition task.
By conducting this process, we can bridge the information gap between the pre-trained ASR task and the target SER task, as confirmed by our experiments in section~\ref{ablation_study}.

As depicted in Figure~\ref{overall} (a), the wav2vec 2.0 model $\mathbf{M}$($\mathbf{W}_{0}$), with pre-trained weights $\mathbf{W}_{0}$, consists of three sub-modules: the feature encoder module, the transformer module, and the quantization module.
%===============================
Specifically, we utilize a CNN-based encoder to encode the $i$-th input unlabeled speech signals into low-dimensional vectors, denoted as $\mathbf{x}_{i}$. Subsequently, we randomly mask 15\% of 
the features (following BERT~\cite{kenton2019bert}) of the speech vectors. We then decode them using two decoders to obtain quantized and context representations. The quantization decoder can transform continuous speech vectors $\mathbf{x}_i$ into discrete codewords from phonemes codebooks~\footnote{A quantized codebook refers to a set of predetermined values or codewords used to represent a continuous signal in a discrete form~\cite{baevski2020wav2vec}.}, resulting in $\mathbf{z}_{i}^{q}$. 
Meanwhile the wav2vec 2.0 decoder (transformer layers) employs self-attention to decode continuous speech vectors $\mathbf{x}_{i}$ into context-aware representations $\mathbf{z}_{i}^{c}$. 
Then, we design contrastive loss~\cite{baevski2020wav2vec} (cl) to minimize the differences between quantized and context representations as follows:
\begin{equation}
\label{contrastive}
    \mathcal{L}_{cl} = -\sum_{i=1}^{n}\text{log}\frac{\text{exp}(\text{sim}(\mathbf{z}_{i}^{c}, \mathbf{z}_{i}^{q})/\kappa)}{\sum_{j=1}^{n}\text{exp}(\text{sim}(\mathbf{z}_{i}^{c}, \mathbf{z}_{j}^{q})/\kappa)},
\end{equation}
where the temperature hyperparameter $\kappa$ is set to 0.1,  and $\text{sim}(\mathbf{a}, \mathbf{b}) = \mathbf{a}^{T}\mathbf{b}/\|\mathbf{a}\|_{2}\|\mathbf{b}\|_{2}$, where $T$ represents the transposition of a vector.
Eq. (\ref{contrastive}) can help obtain better quantized and context representations because two decoders can provide highly heterogeneous contexts for each speech signal~\cite{YouCSCWS20}.

To minimize the information gap between the pre-trained model and downstream SER task, following BERT~\cite{kenton2019bert}, we first randomly mask 15\% of the tokens of each speech signal. We then apply reconstruction loss on the corrupted downstream SER dataset to generate tokens for reconstructing the original data, which can be formulated as follows:
\begin{equation}
    \mathcal{L}_{rl} = - \frac{1}{|N_{m}|}\sum_{i=\text{First masked token}}^{\text{Last masked token}} s^{\text{true}}_{i}\text{log}(s_{i}^{\text{predicted}})
\end{equation}
where $N_{m}$ is the number of masked tokens, $s^{\text{true}}_{i}$ and $s_{i}^{\text{predicted}}$ are the ground-truth and predicted token probability of the $i$-th masked token, respectively.

Finally, we combine contrastive loss and reconstruction loss for the TAPT process as follows:
\begin{equation}
    \mathcal{L}_{TAPT} = \mathcal{L}_{cl} + \mathcal{L}_{rl}.
\end{equation}

Please note that although pseudo-TAPT~\cite{2110-06309} also adopts TAPT, we employ different loss functions. We believe that our method is simpler and more suitable for upstream ASR tasks. Specifically, they invest significant time using K-means to extract frame-level emotional pseudo-labels and continually pre-train their model in a supervised manner by predicting their frame-level emotion pseudo-labels.
However, K-means is sensitive to the initial value and outliers~\cite{ZhangHJQAH20}, making Pseudo-TAPT unstable and computationally expensive.

\begin{algorithm}
% % \scriptsize
% \small
\SetKwData{Left}{left}\SetKwData{This}{this}\SetKwData{Up}{up}
\SetKwFunction{Union}{Union}\SetKwFunction{FindCompress}{FindCompress}
\SetKwInOut{Input}{Input}\SetKwInOut{Output}{output}
\caption{Active Learning based Fine-tuning}
\label{algorithm}
\Input{Unlabeled data $\mathcal{D}_{\textrm{pool}}$, Model $\mathbf{M}(\mathbf{W}_{0})$, Acquisition size $k$, Iterations $\tau$, total number of selected samples $N_{c}$, and Acquisition function $\mbox{ac}()$.}
$\mathbf{M}_{\textrm{TAPT}}$ $(\mathcal{D}_{\textrm{pool}}; \mathbf{W}_{0}')$ $\leftarrow$ Train $\mathbf{M}(\mathbf{W}_{0})$ on $\mathcal{D}_{\textrm{pool}}$\;
$Q_{0} \leftarrow$ Clustering-based initialization from $\mathcal{D}_{\textrm{pool}}$\;
$\mathcal{D}_{\textrm{train}}^{0} \leftarrow Q_{0}$;
$\mathcal{D}_{\textrm{pool}}^{0} \leftarrow \mathcal{D}_{\textrm{pool}} \backslash Q_{0}$ where $|Q_{0}|=1\%N_{s}$\;
% $\mathcal{P}_{\textrm{TAPT}}$ $(x;W_{0}')$ $\leftarrow$ Train $\mathcal{P}$ $(x;W_{0})$ on $\mathcal{D}_{\textrm{pool}}$ \;
$\mathbf{M}_{0}([\mathbf{W}_{0}^{'},\mathbf{W}_{c}]) \leftarrow$ Initialized from $\mathbf{M}_{\textrm{TAPT}}$ $(\mathcal{D}_{\textrm{pool}}; \mathbf{W}_{0}^{'})$\;
$\mathbf{M}_{0}(\mathcal{D}_{\textrm{train}}^{0}; [\mathbf{W}_{0}^{'}, \mathbf{W}_{c}]) \leftarrow$ Train $\mathbf{M}_{0}([\mathbf{W}_{0}^{'}, \mathbf{W}_{c}])$ on $\mathcal{D}_{\textrm{train}}^{0}$\;
\For{i $\leftarrow$ 1 to $\tau$}
{$Q_{i}$ $\leftarrow$
$\mbox{ac}(\mathbf{M}_{i-1},\mathcal{D}_{\textrm{pool}}^{i-1},k)$ 
\quad \textcolor{blue}{$\triangleright$} \textcolor{blue}{Annotating $k$ samples}\;
$\mathcal{D}_{\textrm{train}}^{i}$\,\,=\,\,$\mathcal{D}_{\textrm{train}}^{i-1}$ $\cup$ $Q_{i}$  \,\, \textcolor{blue}{$\triangleright$} \textcolor{blue}{Add labeled samples to $\mathcal{D}_{\textrm{train}}^{i}$}\;
$\mathcal{D}_{\textrm{pool}}^{i} \leftarrow \mathcal{D}_{\textrm{pool}}^{i-1} \backslash Q_{i}$ \quad \textcolor{blue}{$\triangleright$} \textcolor{blue}{Delete samples from $\mathcal{D}_{\textrm{pool}}^{i}$}\;
$\mathbf{M}_{i}(\mathcal{D}_{\textrm{train}}^{i}; [\mathbf{W}_{0}^{'}, \mathbf{W}_{c}]) \leftarrow$ \textcolor{blue}{Train  $\mathbf{M}_{i-1}$ on $\mathcal{D}_{\textrm{train}}^{i}$}\; }
\end{algorithm}

\subsection{Active Learning (AL) based Fine-tuning}

When we finish the TAPT process, we obtain the model $\mathbf{M}_{\textrm{TAPT}}(\mathbf{W}_{0}^{'})$ with $\mathbf{W}_{0}^{'}$ as the weight initialization for the AL process (cf. Line 1 of Algorithm~\ref{algorithm}). 
A typical AL setup starts by treating $\mathcal{D}$ as a pool of unlabeled data $\mathcal{D}_{\textrm{pool}}$ and performs $\tau$ iterations of sample selection. 
Specifically, in the $i$-th iteration, $k$ samples are selected using a given acquisition function $ \mbox{ac}()$: $Q_{i}$ = $\{\mathbf{x}_{1},\cdots,\mathbf{x}_{k}\}$. Here, $k$ is a variable parameter. We determine it based on the number of iterations $\tau$ and the predefined total number of selected samples $N_{s}$, i.e., $k= \text{ROUND}(N_{s}/\tau)$,  where ROUND() rounds the number down.
For example, we adopt Entropy~\cite{roy2001toward} as the $\mbox{ac}()$ function to measure the uncertainty of the samples and select the most uncertain $k$ samples. These selected samples are then labeled and added to the $i$-th training dataset $\mathcal{D}^{i}_{\textrm{train}}$, with which a model is fine-tuned for SER.
%===========

%
One primary goal of \textsc{After} is to explore whether AL strategies can reduce the number of annotation samples, as labeling large-scale datasets is the most laborious part of SER. Instead of focusing on proposing new active learning query strategies, we adopt five of the most well-known and influential AL strategies for evaluation,  including Entropy~\cite{roy2001toward}, Least Confidence~\cite{DredzeC08}, Margin Confidence~\cite{DredzeC08}, ALPs~\cite{YuanLB20}, and BatchBald~\cite{KirschAG19}. These methods use different criteria to help select the most uncertain and informative samples from $\mathcal{D}_{\textrm{pool}}$, {and we introduce them briefly in this paper.

\noindent \ding{172} Entropy measures the uncertainty of $\mathbf{x}_{i}$ as \begin{equation} \textrm{Entropy}(\mathbf{x}_{i}) = -\sum_{j=1}^{c}P(\hat{y}_{j}|\mathbf{x}_{i})\textrm{log}P(\hat{y}_{j}|\mathbf{x}_{i}), \end{equation} where $c$ is the number of emotional classes and $P(\hat{y}_{j}|\mathbf{x}_{i})$ represents the predicted probability of $\mathbf{x}_{i}$ for the $j$-th emotion.
% $\textrm{Entropy}(x_{i})$ = $-\sum_{j=1}^{c}P(y_{j}|x_{i})\textrm{log}P(y_{j}|x_{i})$. 

\noindent \ding{173} Least Confident measures the most incontinent samples as
\begin{equation} \textrm{Least Confident}(\mathbf{x}_{i}) = \sum_{j=1}^{c} (1-P(\hat{y}_{j}|\textbf{x}_{i}))
\end{equation} 
where $c$ is the number of emotional classes and $P(y_{j}|\mathbf{x}_{i})$ represents the predicted probability of $\mathbf{x}_{i}$ for the $j$-th emotion.

\noindent \ding{174} Margin Confidence  is the process of selecting the sample with the smallest difference between the maximum and second largest probability predicted by the model, which can be formulated as
\begin{equation} \textrm{Margin Confidence}(\mathbf{x}_{i}) = (P(\hat{y}_{1}|\textbf{x}_{i}) - P(\hat{y}_{2}|\textbf{x}_{i}))
\end{equation} 
where $P(y_{1}|\mathbf{x}_{i})$ represents the largest predicted probability of $\mathbf{x}_{i}$ and $P(y_{2}|\mathbf{x}_{i})$ represents the second largest predicted probability of $\mathbf{x}_{i}$.

\noindent \ding{175} ALPs iteratively selects the sample closest to the cluster center as the most differentiated and informative sample each batch. This can be formulated as follows:
\begin{equation} \textrm{ALPs}(\mathbf{x}) = \textrm{argmin}_{\textbf{x}_{i}}\|\textbf{centers}-\textbf{x}_{i}\|
\end{equation} 
where $\textbf{centers}$ is the clustering centers (we follow their paper using K-means to obtain clustering centers).

\noindent \ding{176} BatchBald jointly score samples by estimating the mutual information ($\mathbb{I}$) between a set of multiple data points and the model parameters:
\begin{equation} \textrm{BatchBald}
(\{\textbf{x}_{1}, \dots, \textbf{x}_{b}\}, p(\textbf{w}|\mathcal{D}_{\textrm{train}})) = \mathbb{I}(y_{1},\dots,y_{b};\textbf{w}|\textbf{x}_{1},\dots,\textbf{x}_{b},\mathcal{D}_{\textrm{train}}).
\end{equation}

After applying the above query strategies for the samples,
we select the $k$ most uncertain or diverse samples for annotation and add them to the training dataset $\mathcal{D}_{\textrm{train}}$. 
Traditional AL methods often use random initialization; however, we found that these AL methods are sensitive to the initialization process, leading to the selection of redundant samples or outliers in each AL iteration with poor initialization. Therefore, instead of directly using AL methods, we propose a clustering-based initialization for all AL methods (K-means in this study), resulting in better performance (details about K-means are given in section~\ref{Active_learning_results}). 
Please note that, as illustrated in Algorithm~\ref{algorithm}, clustering-based initialization is applied only in the initialization process, and subsequent iterations of the AL loop do not require a K-means process.

\subsection{Initialization for Active Learning}
\label{initialization}

We observe that AL methods are particularly sensitive to initialization,  as the initial set of samples can substantially impact the selection order of subsequent samples in each iteration of AL.
However, most AL methods randomly select 1\% of samples for initialization~\cite{MargatinaVBA21}. In contrast, we propose a novel clustering-based (K-means) initialization method to improve the performance of SER. Specifically, we first extract sample representations of the training data from the wav2vec 2.0 CNN-based encoder. Then, we employ K-means on the training data and select 1\% of samples closest to the cluster centers as our initialized samples. It is important to note that we use the elbow method~\cite{DBLP:journals/cin/SammoudaE21} to automatically determine the number of clusters for K-means, and we use the Euclidean distance to measure the distance between sample representations.

\subsection{Emotion Recognition Classifier}

As shown in Figure.~\ref{overall} (b), we incorporate a task-specific classification layer with additional parameters $\mathbf{W}_{c}$ for emotion recognition on top of wav2vec 2.0. We fine-tune the classification model $\mathbf{M}_{i}([\mathbf{W}_{0}^{'}, \mathbf{W}_{c}])$ in each AL iteration using} all labeled samples in $\mathcal{D}_{\textrm{train}}$ (cf. Lines 6-10 of Algorithm~\ref{algorithm}). We formulate the cross-entropy loss for the emotion recognition classifier as follows:
\begin{equation}
    \mathcal{L}_{ce} = -\frac{1}{k} \sum_{i=1}^{k}\sum_{j=1}^{c} y^{j}_{i} \log{ (\hat{y}_{i}^{j})}, 
\end{equation}
where $c$ is the number of emotion classes,  $k$ is the number of selected samples at $t$-th iteration, $\hat{y}_{i}^{j}$ is the $i$-th predicted label, and $y_{i}^{j}$ is the $i$-th ground-truth of the $j$-th class. 

\section{Experiment Settings}\label{sec:experiment1}

In this section, we first introduce all datasets used in this study in Section~\ref{datasets}. 
Following this, we present the selected baselines in Section~\ref{baselines} and provide implementation details in Section~\ref{implementation}.
Finally, we delve into the detailed active learning strategies used in the following experiments in Section~\ref{Active_learning_results}.

\subsection{Datasets}
\label{datasets}

\noindent \textbf{IEMOCAP:} \quad We first evaluated the performance of all baseline models using the widely used benchmark dataset, IEMOCAP~\cite{BussoBLKMKCLN08}. 
IEMOCAP is a multimodal database commonly employed to evaluate SER performance. It comprises five conversation sessions, each featuring a female and a male actor engaging in improvised and scripted scenarios. The dataset includes 10,039 speech utterances, all sampled at 16kHz with a 16 bits resolution.
To ensure a fair comparison, we merged the ``excited'' class into the ``happy'' class, resulting in four considered emotions: neutral, happy, angry, and sad. Following the approach of \cite{2110-06309}, we adopted a 5-fold cross-validation method, where each IEMOCAP session served as the test set. Additionally, we randomly selected 10\% of the data from the remaining four sessions for our validation dataset, with the rest allocated to our training dataset.\\

%%%%%%%%%%%%%

\noindent \textbf{SAVEE:} \quad To explore the performance of \textsc{After} with broader range of emotions, we incorporated an additional datasets, the Surrey Audio-Visual Expressed Emotion
(SAVEE) dataset~\cite{jackson2014surrey}. SAVEE contains four male speakers: DC, JE, JK, and KL. Each speaker reads the same set of 120 sentences, labeled with one of seven emotion categories: angry, disgust, sad, fear, happy, surprise, and neutral. Utilizing all emotion categories, the dataset comprises 480 utterances, totaling 30 minutes of speech. For fair comparisons with SOTA approaches in experiments, following the previous works~\cite{tuncer2021automated,ye2022gm,wen2022ctl,farooq2020impact,ye2023temporal}, we mainly conducted 10-
fold cross-validation. In each fold, we allocated 90\% of the data for training and 10\% for testing to evaluate the model's fitting ability.\\

\noindent \textbf{Merged dataset:} \quad Many existing methods are inadequate for real-world applications and are susceptible to noise due to their heavy reliance on fine-tuning models using specific small-scale datasets. For example, pseudo-TAPT~\cite{2110-06309} is fine-tuned by the training dataset of IEMOCAP and performs well on IEMOCAP. However, pseudo-TAPT performs poorly when tested on other datasets. 
To provide a potential solution to address this issue, we conducted additional experiments by training on two larger noisy and heterogeneous datasets. We achieved this by merging various datasets from different sources to simulate the noisy environments encountered in real-world scenarios. It is important to note that any emotion recognition datasets containing the corresponding emotional categories can be incorporated into the Merged datasets. In this draft, we selected five widely-used datasets with different languages, recording equipment, and number of actors.
We first introduce each dataset of the Merged dataset as follows:
\begin{itemize}
    \item EmoDB~\cite{BurkhardtPRSW05} database is a freely available German emotional database, created by the Technical University, Berlin, Germany. It features ten professional speakers, including five males and five females, who participated in the data recording process. The database contains a total of 535 utterances and comprises seven emotions: anger, boredom, anxiety, happiness, sadness, disgust, and neutral. The data was recorded at a 48-kHz sampling rate and then down-sampled to 16-kHz.
    \item ShEMO~\cite{MohamadNezami2019} database includes 3,000 semi-natural utterances, totaling three hours and 25 minutes of speech data extracted from online radio plays. ShEMO encompasses speech samples of 87 native-Persian speakers, covering six basic emotions: anger, fear, happiness, sadness, surprise, and neutral states.
    \item RAVDESS~\cite{0196391} database contains 7,356 files and features 24 professional actors (12 female, 12 male). Each actor vocalizes two lexical-matched statements in a neutral North American accent. The speech includes calm, happy, sad, angry, fearful, surprise, and disgust. Each expression is produced at two levels of emotional intensity (normal, strong), alongside an additional neutral expression. All conditions are available in three modality formats: Audio-only, Audio-Video, and Video-only.
    \item EMov-DB~\cite{adigwe2018emotional} database includes recordings from four speakers, including two males and two females. The emotional styles covered include neutral, sleepiness, anger, disgust, and amused. Each audio file is recorded in 16bits .wav format.
    \item CREMA-D~\cite{CaoCKGNV14}  database is an emotional multimodal actor dataset consisting of 7,442 original clips from 91 actors. These clips were from 48 male and 43 female actors ranging in age from 20 to 74, representing a variety of races and ethnicities. Actors speak from a selection of 12 sentences, each presented with one of six different emotions: anger, disgust, fear, happy, neutral, and sad, and across four different emotion levels: low, medium, high, and unspecified.
\end{itemize}

\begin{table}[h]
\caption{Descriptive statistics of the Merged dataset, which contains speeches in three types of languages. The ratio of the four labels is in the order of Anger : Neutral : Sad : Happy.}
\centering
\setlength{\tabcolsep}{3mm}{
\begin{tabular}{@{}lccc@{}}
\toprule
\multirow{2}{*}{\textit{\textbf{Datasets}}} & \multicolumn{3}{c}{\textit{\textbf{Characteristics}}} \\ \cmidrule(l){2-4} 
& \# Samples & \# Actors & Ratio of Four Labels  \\
\midrule 
\midrule
IEMOCAP~\cite{BussoBLKMKCLN08} (English) & 10,038 & 2 & \,2.5\,:\,1.2\,:\,2.4\,:\,1.0  \\
 EmoDB~\cite{BurkhardtPRSW05} (German) & 408 & 10 & \,3.1\,:\,1.3\,:\,1.0\,:\,1.1   \\
 ShEMO~\cite{MohamadNezami2019} (Persian) & 2,737 & 87 & \,5.3\,:\,5.1\,:\,2.2\,:\,1.0  \\
 RAVDESS~\cite{0196391} (English) & 672 & 24 & \,2.0\,:\,1.0\,:\,2.0\,:\,2.0  \\ 
 EMov-DB~\cite{adigwe2018emotional} (English) & 3,038 & 4 & \,1.4\,:\,1.0\,:\,0.0\,:\,0.0  \\ 
 CREMA-D~\cite{CaoCKGNV14} (English) & 4,900 & 91 & \,1.0\,:\,1.7\,:\,1.0\,:\,1.0  \\ 
\midrule
\textbf{Merged dataset} & 21,793 & 218  & \,1.5\,:\,1.4\,:\,1.0\,:\,1.5  \\
\bottomrule
\end{tabular}}\label{table1}
\end{table}

As shown in Table~\ref{table1}, we manually controlled the number of instances for each of the four labels in the Merged dataset to maintain label balance.
Different from IEMOCAP, EmoDB is a German emotional database, ShEMO is a Persian emotional database, and both RAVDESS and CREMA-D contain more actors (24 actors and 91 actors, respectively). 
We constructed the Merged dataset by merging the training data of the mentioned following datasets with the training data of IEMOCAP.
To explore whether the Merged dataset could improve performance on a single dataset, such as IEMOCAP, we employed a 5-fold cross-validation approach. This involved leaving each IEMOCAP session out as the test set and randomly selecting 10\% of the dataset from the remaining Merged dataset as our validation dataset, while the remainder was allocated for training purposes.
It is important to note that we only use the training data for both the TAPT and AL-based fine-tuning processes to prevent data leakage during evaluation. Furthermore, the training procedures are conducted from scratch separately for the IEMOCAP, SAVEE, Merged, Merged-2, and Merged-3 dataset.\\

\noindent \textbf{Merged-2 dataset:} \quad Merged dataset contains almost acted speech datasets. To better simulate ``real-world'' scenarios, we incorporated two additional spontaneous datasets to Merged dataset to construct Merged-2 dataset: AFEW5.0 and BAUM-1s.
The two added datasets are as follows:
\begin{itemize}
    \item AFEW5.0~\cite{10.1145/2818346.2829994} is a spontaneous audiovisual emotional video dataset developed for emotion recognition in the wild (EmotiW) challenge in 2015. It contains seven emotional categories: anger, disgust, fear, joy, neutral, sadness, and surprise. These emotions were annotated by 3 annotators. The dataset is divided into three parts: train set (723 samples), validation (val) set (383 samples), and test set (539 samples). In this work, we used the train and val sets to validate the performance of our method since the Test set is only available to participants in competitions.
    \item BAUM-1s~\cite{7451244} is a spontaneous audiovisual affective face database of affective and mental states developed in 2016, featuring 31
    Turkish individuals. The video samples were collected in real scenarios, where emotions were elicited by watching films in an unscripted and unguided way. The target emotions include seven basic ones: joy, anger, sadness, disgust, fear, neutral, and surprise, as well as boredom and contempt. Several mental states, such as unsure (confused, and undecided), thinking, concentrating, and bothered, are also included.
\end{itemize}

We only used four emotion categories from AFEW5.0 and BAUM-1s, including anger, neutral, sad, and happy, to construct the Merged-2 dataset. And we used the test data of IEMOCAP as test data for the Merged-2 dataset. Training and evaluation processes are similar to those of the Merged dataset. \\

\noindent \textbf{Merged-3 dataset:} \quad To demonstrate the performance of \textsc{After} on spontaneous datasets, we used the test set of BAUM-1s, which comprises seven emotion categories, for evaluation. The detailed implantation for evaluation follows the approach outlined in \cite{DBLP:journals/taffco/ZhangZT22}.
Additionally, we constructed the Merged-3 dataset by merging two acted speech datasets (EmoDB, RAVDESS) with two spontaneous datasets (AFEW5.0, BAUM-1s).

\subsection{Baselines}
\label{baselines}

We selected different SOTA baselines for different datasets. 
For the IEMOCAP dataset, Merged dataset, and Merged-3 dataset, we selected the best-performing methods:  LSSED~\cite{9414542}, GLAM~\cite{9747517}, RH-emo~\cite{abs-2204-02385}, Light~\cite{Aftab}, Pseudo-TAPT~\cite{2110-06309}, and w2v2-L-robust~\cite{10089511}. 
For the SAVEE dataset, we selected the recently best-performing approaches: DCNN~\cite{farooq2020impact}, TSP+INCA~\cite{tuncer2021automated}, CPAC~\cite{wen2022ctl}, GM-TCN~\cite{ye2022gm}, and TIM-Net~\cite{ye2023temporal}.
For the Merged-3 dataset, we select the baselines as MFCC+PLP+SVM~\cite{7451244}, CNN+SVM~\cite{DBLP:journals/tcsv/ZhangZHGT18}, CNN+DTPM+SVM~\cite{DBLP:journals/tmm/ZhangZHG18}, CNN+SVM~\cite{MA2019184} and CNN+LSTM~\cite{DBLP:journals/taffco/ZhangZT22}.

\subsection{Implementation details}
\label{implementation}

All experiments used the same learning rate of $10^{-4}$ with the Adam optimizer. 
Our implementation of wav2vec 2.0 (wav2vec2-base) is based on the Hugging Face framework~\footnote{\url{https://huggingface.co/facebook/wav2vec2-base}}. The audio window length was set to 20 ms. 
We fine-tuned the model in a few-shot manner, involving longer fine-tuning, more evaluation steps during training, and early stopping with 20 epochs based on validation loss.
%========================
To ensure fair comparison with previous studies, we employ either off-the-shelf software packages or utilize the provided code by respective authors. Each model underwent ten executions, and the average performance across these runs is considered the final result. The hyper-parameters are chosen as default if provided, or tuned otherwise.
Following the work~\cite{10095777}, we evaluated the models using weighted accuracy (WA) and unweighted accuracy (UA)~\cite{metallinou2010decision} in speaker-independent settings. Please note that we did not require the data to be labeled by actual annotators. Instead,  we used the ground-truth labels available in the training dataset. Specifically, we masked the labels and only receive them when the AL methods determined that the samples should be labeled. This approach is a common technique used by AL researchers to validate their methods~\cite{roy2001toward}. However, it is worth mentioning that in real-world scenarios,
human annotators would be responsible for labelling the data.

\subsection{Active Learning Strategies Selection for \textsc{After}}
\label{Active_learning_results}

As shown in Figure~\ref{overall} (c), \textsc{After} incorporates an AL strategy for sample selection. To identify the most suitable AL method for \textsc{After}, we combined it with multiple well-known AL methods and evaluated their performance. Furthermore, we find that AL methods are sensitive to initialization, with most AL methods randomly selecting 1\% samples for initialization~\cite{MargatinaVBA21}. Unlike them, we proposed a novel clustering-based (K-means) initialization method to improve the performance of SER.
Specifically, we first extract sample representations of training data from the wav2vec 2.0 CNN-based encoder. Then, we employed K-means on the training data and selected 1\% of samples closest to the cluster centers as our initialized samples. Please note that we use the elbow method~\cite{DBLP:journals/cin/SammoudaE21} to determine the number of clusters for K-means automatically, and we use the Euclidean distance to measure the distance between sample representations.

\begin{figure*}[h]
\centering
\includegraphics[width=0.8\textwidth]{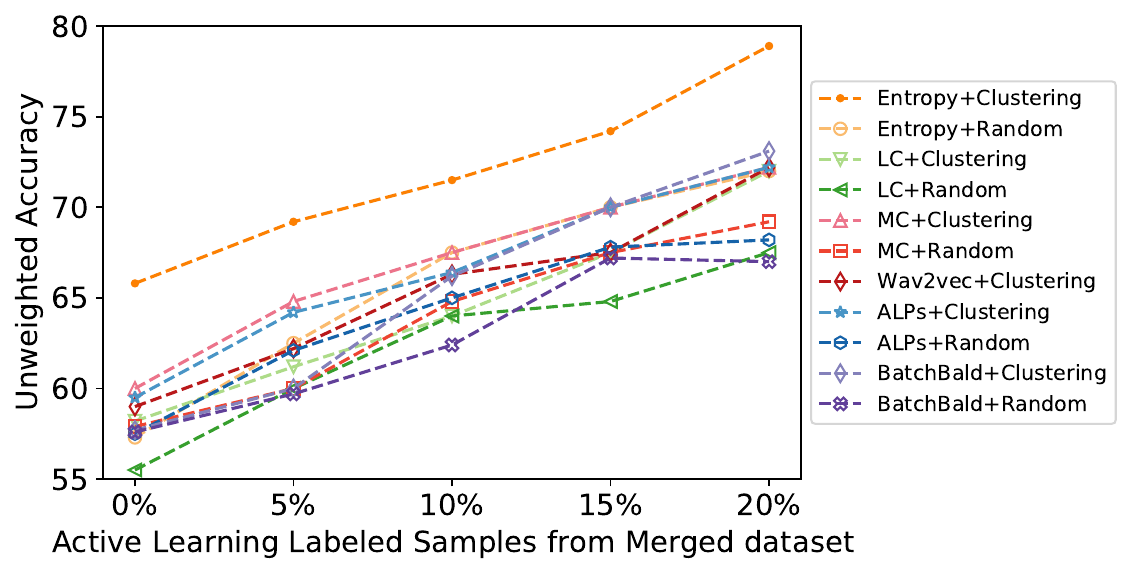}
\caption{Ratio of labeled samples vs. Unweighted Accuracy.}
\label{Ratio}
\end{figure*}

Figure~\ref{Ratio} shows that clustering-based initialization outperformed the random initialization for all AL methods. 
The initial set of samples significantly influenced the selection order of samples in each iteration of AL, and an effective initialization enhanced the performance and stability of AL methods. 
Figure~\ref{Ratio} illustrates that \textit{Entropy+Clustering} emerged as the most effective AL strategy for \textsc{After} on the Merge dataset. 
Although we only displayed the diagram for UA due to space constraints, the diagram for WA exhibited similar trends.
\textbf{Therefore, \textit{Entropy+Clustering} was selected as the primary AL method for \textsc{After}}. 
We recommend using \textit{Entropy+Clustering}, the simplest yet most efficient strategy for real-world applications.

\begin{table}[h]
\caption{After with Entropy to select 10\%-100\% labeled samples of the Merge dataset for fine-tuning.}
\centering
\setlength{\tabcolsep}{7mm}{
\begin{tabular}{@{}lcccccc@{}}
\toprule
\multirow{1}{*}{\textit{\textbf{Datasets}}} & \multicolumn{6}{c}{\textit{\textbf{AFTER (TAPT+ AL-based FT)}}} \\ 
%\cmidrule(l){2-7} 
\toprule
& 10\% & 20\%  & 40\%  & 60\%  & 80\%  & 100\%   \\
\midrule
\midrule
\textbf{UA}  & 71.45 & \textbf{77.41}  & 78.64 & \textbf{79.32} & 79.26 &79.15 \\
\textbf{WA}  & 69.01 & \textbf{74.32}  & 75.48 & \textbf{76.03} & 75.92 &75.94 \\
\textbf{Time} (mins) & 262.8 & \textbf{316.4}  & 785.4 & \textbf{942.2} & 1182.6 & 1508.2\\
\bottomrule
\end{tabular}}\label{tab:tablenew}
\end{table}

We analyzed the relationship between the ratio of labeled samples, performance, and time consumption of \textsc{After}. Results in Table~\ref{tab:tablenew} show that both performance and time consumption  of \textsc{After} increased as the ratio of labeled samples increased. \textbf{Our findings indicate that using 20\% labeled samples yielded a significant improvement in performance while reducing the time consumption by \underline{79\%} compared to fine-tuning on 100\% samples.} Thus, we selected 20\% labeled samples as a trade-off between performance and time consumption for subsequent experiments.

\section{Experimental Results and Discussion}\label{sec:experiment2}

\subsection{Comparison with Other Initialized Strategies}
\label{initialized-methods}

As illustrated in Section~\ref{initialization}, we propose a simply but efficient K-means initialization method 
applicable to various SER tasks. While representative-based methods like BMAL~\cite{DBLP:journals/pami/ChakrabortyBSPY15} and density-based methods like DACS~\cite{10.1145/3534678.3539476} are also available, our focus remains on demonstrating the effectiveness of our proposed initialization method.
To assess the effectiveness of our approach, we compare it with BMAL, DACS, and random sampling as baseline initialization methods. Specifically, the selected initialization baselines are introduced as follows.

DACS~\cite{10.1145/3534678.3539476} is a density-aware Core-set approach used to estimate sample densities and selectively choose diverse points from sparse regions. For each input $\bm{x}_{i}$, the density score for each sample is calculated as follows:
\begin{equation}
    \text{Density}(\bm{x}_{i}) = \frac{1}{k} \sum_{j\in \mathcal{N}(\bm{x}_{i}, k)} \|\bm{x}_{i}-\bm{x}_{j}\|_{2}^{2},
\end{equation}
where $\mathcal{N}(\bm{x}_{i},k)$ represents the $k$-nearest neighbors of $\bm{x}_{i}$~\cite{10.1145/3534678.3539476}. We use default parameters from their original paper, selecting
the top 1\% of samples with the highest scores as initialized samples for downstream tasks.

BMAL~\cite{DBLP:journals/pami/ChakrabortyBSPY15} considers the distance between a sample and its surrounding labeled samples to enrich the diversity of the labeled dataset. Diversity is measured by the KL-divergence of the class probabilities distribution of similar neighboring instances, formulated as:
\begin{equation}
    \text{Divergence}(\bm{x}_{i},\bm{x}_{j})= \sum_{j} P(\hat{y}_{j}|\bm{x}_{i})- P(\hat{y}_{j}|\bm{x}_{j})\log \frac{P(\hat{y}_{j}|\bm{x}_{i})}{P(\hat{y}_{j}|\bm{x}_{j})},
\end{equation}
where $\hat{y}_{j}$ is the predicted label for the $j$-th sample. We use the default parameters from their original paper, selecting the top 1\% of samples with the highest scores as initialized samples for downstream tasks.
% We use Entropy as our active learning strategy and set the initialization size as 1\% of the training samples. For our experiments, We use HuBERT (hubert-base-ls960) form Hugging Face~\footnote{https://huggingface.co/facebook/hubert-base-ls960}.} 

\begin{figure*}[h]
\centering
\includegraphics[width=0.8\textwidth]{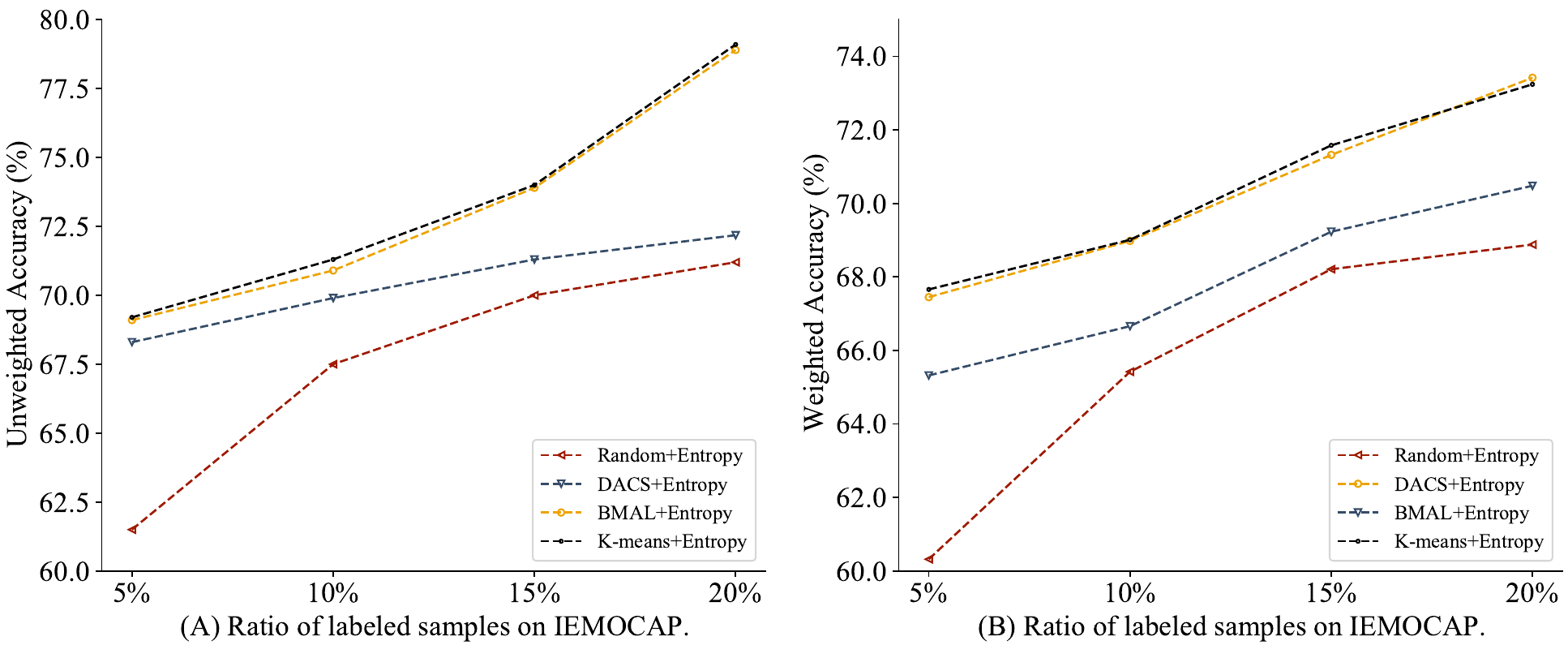}
\caption{Comparison of various initialization methods for AL, with Entropy employed as the active learning strategy. Initialization involves selecting 1\% of the samples.}
\label{figure:initialized}
\end{figure*}

As depicted in Figure~\ref{figure:initialized}, we observe that K-means exhibits comparable performance to DACS and outperforms BMAL and Random Sampling. 
DACS operates as a density-aware core-set approach.
When selecting 1\% diverse samples for initialization, both K-means and DACS tend to choose the same sample set nearest to each clustering center. In contrast, BMAL and Random Sampling 
struggle to select the most representative samples, leading to decreased classification performance.
Instead of employing DACS with its numerous hyperparameters requiring adjustment, we opted for the simple yet efficient K-means as our initialization method. 
Furthermore, our future work will explore the utilization of more representative-based methods for initialization.

\begin{table}[h]
\centering
\caption{Overall performance comparison on 4 emotion categories. 
\textsc{After} adopted Entropy+Clustering and selected 20\% samples for fine-tuning. Baselines use all samples from each corresponding training data for training. The symbol $\dag$ indicates that \textsc{After} significantly surpassed all baselines with $p<0.05$ according to the t-test.}
\setlength{\tabcolsep}{5mm}{
\begin{tabular}{@{}lcccccc@{}}
\toprule
\multicolumn{1}{l}{\multirow{2}{*}{\textbf{\textit{Methods}}}} 
& \multicolumn{2}{c}{\textbf{\small{IEMOCAP}}} 
& \multicolumn{2}{c}{\textbf{\small{Merged Dataset}}} 
& \multicolumn{2}{c}{\textbf{\small{Merged-2 Dataset}}}
\\   
% & \multicolumn{2}{c}{\textbf{\small{Dublin}}}    
% & \multicolumn{2}{c}{\textbf{\small{arXiv}}}  \\ 
\cmidrule(lr){2-3} \cmidrule(lr){4-5} \cmidrule(lr){6-7} 
% \cmidrule(lr){6-7} \cmidrule(lr){8-9}   
\multicolumn{1}{c}{}                         & UA $\uparrow$	         & WA  $\uparrow$       & UA  $\uparrow$        & WA $\uparrow$          & UA  $\uparrow$        & WA $\uparrow$           \\
\midrule
\midrule
GLAM [2022]                   & 74.01          & 72.98          & 71.38          & 69.28     & 70.21   & 68.32                \\
LSSED [2021]                    & 73.09          & 68.35          & 25.00          & 36.20    & 22.35 & 33.47                      \\
RH-emo [2022]                    & 68.26          & 67.35          & 43.20          & 42.80      & 42.18 &40.78                    \\
Light [2022]                    & 70.76          & 70.23          & 69.28          & 71.38      &68.36 &70.27                    \\
Pseudo-TAPT [2022]                    & 74.30          & 70.26          & 71.25          & 68.83   &70.38 &68.74    \\
w2v2-L-robust [2023] &74.28 &70.23 &71.22 &68.77 & 71.64 & 68.98\\
\midrule
\textbf{\textsc{After}}                                      & $ \,\,\textbf{76.07}^{\dag}$ & $ \,\,\textbf{73.24}^{\dag}$  & $ \,\,\textbf{77.41}^{\dag}$ &$ \,\,\textbf{74.32}^{\dag}$ &$ \,\,\textbf{77.59}^{\dag}$ &$ \,\,\textbf{74.38}^{\dag}$  \\
\bottomrule
\end{tabular}
}\label{performance}
\end{table}

\begin{table}[h]
\centering
\caption{Comparison of baseline architectures. All information is reported by summarizing from their original papers.}
\setlength{\tabcolsep}{6mm}{
\begin{tabular}{@{}lccc@{}}
\toprule
Methods & Backbone & Backbone Size & Pre-training Datasets 
\\   
\midrule
\midrule
GLAM [2022]                   & CNNs          & 15 M          & Without pre-training                         \\
LSSED [2021]                    & ResNet152          & 60 M          &   ImageNet-1k                            \\
RH-emo [2022]                    & ResNet50	          & 25 M          &  ImageNet-1k and IEMOCAP                             \\
Light [2022]                    & CNNs          & 7 M          & Without pre-training                            \\
Pseudo-TAPT [2022]                    & wav2vec 2.0 Base          & 94 M          &  Librispeech~\footnote{https://www.openslr.org/12} and IEMOCAP            \\
w2v2-L-robust [2023] & wav2vec 2.0 Large &317 M & Librispeech and Libri-Light~\footnote{https://github.com/facebookresearch/libri-light} \\ 
\midrule
\textbf{\textsc{After}}                                      & wav2vec 2.0 Base & 94 M  & Librispeech and IEMOCAP   \\
\bottomrule
\end{tabular}
}\label{parameters_baselines}
\end{table}

%%%%%%%%%%%%%%%%%%%%%%%%%%%

\subsection{Comparison with Best-performing Baselines}
\label{sec.compare with sota}

Table~\ref{performance} displays the main results of \textsc{After} and the baseline methods on three datasets (IEMOCAP, Merged dataset, and Merged-2 dataset) in terms of UA and WA. 
To provide a more detailed comparison of baseline performance,
we outline the backbone architecture, backbone size, and pre-training datasets of the baselines in Table~\ref{parameters_baselines}.
\textsc{After} demonstrated superior performance compared to all baselines, achieving this with only 20\% labeled samples for fine-tuning, whereas existing baselines use the entire datasets for training.
Specifically, on the IEMOCAP dataset, \textsc{After} improved UA and WA by \textbf{2.38\%} and \textbf{0.36\%}, respectively, compared to the SOTA baseline (UA of Pesudo-TAPT and WA of GLAM). 
Furthermore, on the Merged dataset, \textsc{After} improved UA and WA by \textbf{\underline{8.45\%}} and \textbf{4.12\%}, respectively, compared to the SOTA baseline (UA of GLAM and WA of Light).
Regarding the Merged-2 dataset, \textsc{After} showed improvements of \textbf{8.30\%} and \textbf{5.84\%} in UA and WA, respectively, compared to UA of w2vw-L-r-12 and WA of Light.

Based on Tables~\ref{performance} and \ref{parameters_baselines}, we have four findings to share as follows: 
\begin{itemize}
    \item (1) \textbf{Larger-scale pre-training models yield better performance:} 
    Traditional CNN-based backbones were insensitive to the pre-training process, as evidenced by GLAM (without pre-training) outperforming LSSED and RH-emo (pre-trained with ResNet). Conversely, larger-scale wav2vec 2.0-based pre-training methods, such as pseudo-TAPT and w2v2-L-robust, significantly outperformed CNN-based models, benefiting from a broader range of hyperparameters and larger pre-training datasets. Even when employing the same backbone as pseudo-TAPT, our method \textsc{After} outperformed pseudo-TAPT and w2v2-L-robust on all three datasets, demonstrating the effectiveness and applicability of active learning for SER.
    \item (2) \textbf{Larger-scale pre-training models exhibit denoising capabilities to a certain extent:} wav2vec 2.0-based methods significantly outperformed CNN-based methods on the Merged dataset and Merged-2 dataset, while showing comparable performance on the IEMOCAP dataset. LSSED~\cite{9414542} and RH-emo~\cite{abs-2204-02385} achieved favorable results with IEMOCAP but showed poor performance with the Merged dataset and Merged-2 dataset, possibly due to their limited denoising and domain transfer capabilities. In contrast, GLAM~\cite{9747517} and Light~\cite{Aftab} employ multi-scale feature representations and deep convolution blocks to capture high-level global data features, advantageous for filtering out noisy low-level features and enhancing performance across all datasets. Pseudo-TAPT~\cite{2110-06309}, w2v2-L-robust, and \textsc{After} adopt larger-scale pre-training models, which understand relevant features for the downstream SER task and help denoise irrelevant or noisy features to improve robustness against real-world datasets.
    \item (3) \textbf{Active Learning can help achieve better performance on real-world datasets:} \textsc{After} achieved superior classification accuracy when utilizing the Merged dataset and Merged-2 dataset compared to solely relying on the IEMOCAP. However, baselines achieved their optimal performance solely with the IEMOCAP, as they are susceptible to the influence of outliers and redundant data. Pseudo-TAPT~\cite{2110-06309} enhances model robustness by using K-means to capture higher-level frame emotion labels as pseudo labels for supervised TAPT. Although baselines can mitigate dataset noise to a certain extent, they exhibit high time complexity during fine-tuning with large-scale datasets and fail to effectively bridge the gap between pre-training and the downstream SER task. In contrast, \textsc{After} uses unsupervised TAPT to mitigate the information gap between the source domain (ASR) and the target (SER) domain. Additionally, \textsc{After} selects a subset of the most informative and diverse samples for iterative fine-tuning, offering three advantages: Firstly, it reduces labor consumption for manually labeling large-scale SER samples; Secondly, by utilizing a smaller labeled dataset, \textsc{After} significantly reduces the overall time consumption (Figure~\ref{time}), making it practical and feasible for real-world applications; Finally, the iterative fine-tuning process employed by \textsc{After} improves performance and stability by eliminating noise and outliers present in the selected samples, leading to enhanced overall model performance in SER tasks.
    \item (4) \textbf{Baselines performed better on the Merged-2 dataset than on the Merged-1 dataset:} \textsc{After} achieved superior classification performance on the Merged-2 dataset compared to the Merged dataset. However, the combination of acted and spontaneous speech datasets posed greater challenges for other baselines due to their sensitivity to heterogeneous and noisy samples.
    Merging multiple datasets enabled \textsc{After} to extract a wider variety of samples from a larger pool of data. 
    This increased diversity of samples contains more information, 
    resulting in improved classification performance.
\end{itemize}

\begin{table}[h]
\centering
\caption{Overall performance comparison on the SAVEE dataset with seven emotion categories. \textsc{After} adopted Entropy+Clustering and selected 20\% samples for fine-tuning. We obtained the baselines' performance directly from TIM-Net.}
\setlength{\tabcolsep}{27mm}{
\begin{tabular}{@{}lcc@{}}
\toprule
\textbf{\textit{Methods}}                     
& UA $\uparrow$	         & WA  $\uparrow$                     \\
\midrule
\midrule
DCNN [2020]                   & -          & 82.10                            \\
TSP+INCA [2021]   & 83.38          & 84.79      \\
CPAC [2022]                    & 83.69          & 85.63                                    \\
GM-TCN [2022]                    & 83.88          & 86.02                                  \\
TIM-Net [2023]                    & 86.07          & 87.71              \\
\midrule
\textbf{\textsc{After}}                                      & \textbf{86.23} & \textbf{87.98}    \\
\bottomrule
\end{tabular}
}\label{performance-7}
\end{table}

\begin{table}[h]
\centering
\caption{Weighted Accuracy comparison on seven emotion categories. 
\textsc{After} adopted Entropy+Clustering and selected 20\% samples for fine-tuning. Baselines use all samples from each corresponding dataset for training.}
\setlength{\tabcolsep}{12.3mm}{
\begin{tabular}{@{}lcc@{}}
\toprule
\textbf{\textit{Methods}}
& \textbf{\small{BAUM-1s}}
& 
\textbf{\small{Merged-3 Dataset}}
\\   
% & \multicolumn{2}{c}{\textbf{\small{Dublin}}}    
% & \multicolumn{2}{c}{\textbf{\small{arXiv}}}  \\ 
% \cmidrule(lr){2-3} \cmidrule(lr){4-5} 
% % \cmidrule(lr){6-7} \cmidrule(lr){8-9}   
% \multicolumn{1}{c}{}                         & UA $\uparrow$	         & WA  $\uparrow$       & UA  $\uparrow$        & WA $\uparrow$                    \\
\midrule
\midrule
MFCC+PLP+SVM~\cite{7451244}                   & 29.41          & 28.54                               \\
CNN+SVM~\cite{DBLP:journals/tcsv/ZhangZHGT18}                    & 42.28          & 40.68                                  \\
CNN+DTPM+SVM~\cite{DBLP:journals/tmm/ZhangZHG18}                    & 44.67          & 42.33                                  \\
CNN+SVM~\cite{MA2019184}                   & 42.39          & 41.79                                \\
CNN+LSTM~\cite{DBLP:journals/taffco/ZhangZT22}                   & 50.22          & 48.39           \\
\midrule
\textbf{\textsc{After}}                                      & \textbf{50.64} & \textbf{51.24}    \\
\bottomrule
\end{tabular}
}
\label{performance22}
\end{table}

As depicted in Table~\ref{performance-7},
\textsc{After} demonstrated superior classification performance on the SAVEE dataset with seven emotion categories, underscoring its capacity to recognize a wider spectrum of emotions. Specifically, \textsc{After} improved UA and WA by \textbf{0.19\%} and \textbf{0.31\%}, respectively, using only 20\% of samples. By iteratively extracting the most informative and uncertain samples, \textsc{After} fine-tunes the SSL model wav2vec 2.0, effectively removing irrelevant samples and outliers, thereby improving classification performance.

As shown in Table~\ref{performance22}, we also assess the performance of \textsc{After} on BAUM-1s and Merged-3 Dataset (containing spontaneous datasets) with seven emotion categories (Evaluation on BAUM-1s~\cite{DBLP:journals/taffco/ZhangZT22}). We observed that \textsc{After} can achieve SOTA performance on both BAUM-1s and Merged-3 Dataset. \textsc{After} obtained better performance on the Merged-3 dataset by selecting more diverse samples from a large pool of datasets. Baselines lacked the ability to remove noisy data from the merged dataset, decreasing their performance on real-world scenarios.

\subsection{Ablation Study for \textsc{After}}
\label{ablation_study}

We performed an additional ablation study to assess the efficacy of \textsc{After}, as shown in Table~\ref{ablation_studys}. 
Specifically, we conducted fine-tuning (FT) and TAPT+FT on random sample selection and AL-based (Entropy) sample selection with varying ratios of labeled samples, ranging from 10\% to 100\%. To ensure a fair comparison between them, we adopted the same $K$-means initialization for them and other hyperparameters, such as learning rate and random seeds.

\begin{table}[h]
\centering
\caption{Ablation study on the Merged dataset. FT means fine-tuning, and TAPT+FT indicates the adopting of TAPT followed by fine-tuning with the corresponding selected labeled samples. \textsc{After} adopted Entropy to select samples for fine-tuning. Random Sampling and Entropy Sampling utilize the same K-means initialization.}
\setlength{\tabcolsep}{5mm}{
\begin{tabular}{@{}lcccccccc@{}}
\toprule
\multicolumn{1}{l}{\multirow{3}{*}{\textbf{\textit{Methods}}}} 
& \multicolumn{4}{c}{\textbf{\small{Random Sampling}}} 
& \multicolumn{4}{c}{\textbf{\small{Entropy Sampling}}} \\   
% & \multicolumn{2}{c}{\textbf{\small{Dublin}}}    
% & \multicolumn{2}{c}{\textbf{\small{arXiv}}}  \\ 
\cmidrule(lr){2-5} \cmidrule{6-9} 
% \cmidrule(lr){6-7} \cmidrule(lr){8-9}   
\multicolumn{1}{c}{}                        & \multicolumn{2}{c}{\textbf{\small{FT}}}	         & \multicolumn{2}{c}{\textbf{\small{TAPT+FT}}}      &\multicolumn{2}{c}{\textbf{\small{FT}}}        & \multicolumn{2}{c}{\textbf{\small{\textsc{After}}}}                   \\
\cmidrule(lr){2-3} \cmidrule(lr){4-5}  \cmidrule(lr){6-7} \cmidrule{8-9}
\multicolumn{1}{c}{} & UA $\uparrow$ & WA $\uparrow$ & UA $\uparrow$ & WA $\uparrow$  & UA $\uparrow$ & WA $\uparrow$  & UA $\uparrow$ & WA $\uparrow$ \\
\midrule
\midrule
10\% & 50.82 & 48.96 & 70.21 & 68.85 & 68.21 & 66.32 & 71.45 & 69.01 \\
20\%          & 51.37 & 49.92 & 73.82 & 71.33
& 71.07   & 68.21   & \textbf{77.41}   & \textbf{74.32} \\
30\%         & 52.37 & 50.18 &74.49 & 71.89      &72.35 & 69.21 & 78.20 & 75.16                \\
40\%         & 55.68 & 52.21 & 76.01 & 72.28           &73.55 &70.18 & 78.64 & 75.48               \\
60\%         & 60.39 & 59.32 & 77.21 & 74.58          &74.28 & 71.35 & \textbf{79.32} & \textbf{76.03}              \\
80\%         & 58.34 & 56.72 & 78.88 & 75.82          &73.52 & 70.39 & 79.26 & 75.92               \\
100\%                    & 57.21          & 54.12          & 78.21          & 75.36     & 73.89  & 70.89  & 79.15 & 75.94                    \\
\bottomrule
\end{tabular}}\label{ablation_studys}
\end{table}

From Table~\ref{ablation_studys}, we have four interesting observations: (1) Fine-tuning with active learning significantly improved performance compared to random sampling (FT+Entropy vs. FT+Random), regardless of the number of labeled samples. This result demonstrates that the AL-based fine-tuning strategy efficiently eliminates noise and outliers and selects the most informative and diverse samples for fine-tuning; (2) TAPT+FT outperformed FT on both random sampling and Entropy sampling, indicating that TAPT can effectively minimize the domain difference and significantly enhance the performance of the downstream SER task; (3) With the same number of labeled samples, \textsc{After} obtained better results than TAPT+FT+Random on the Merged dataset. However, \textsc{After} with 20\% labeled samples performs worse than TAPT+FT+Random with 80\%$\sim$100\% labeled samples. The reason is that TAPT+FT uses more labeled data for fine-tuning to prevent the model from overfitting and improve its robustness. In a fair comparison with the same size of the training data for fine-tuning, TAPT+FT+Random with 20\% labeled samples performed worse than \textsc{After}(20\%), demonstrating the effectiveness of \textsc{After}; (4) When 100\% of samples are used, AL-based methods significantly outperforms the random sampling method (FT+Random vs FT+Entropy). The main reason is that Random sampling is affected by noise data, and the model constantly corrects the classification boundary, making it difficult to improve the results. Entropy sampling avoids the effect of noisy data by selecting the most informative and diverse samples for FT in advance to fix the classification boundary properly.

\begin{figure*}[h]
\centering
\includegraphics[width=1\textwidth]{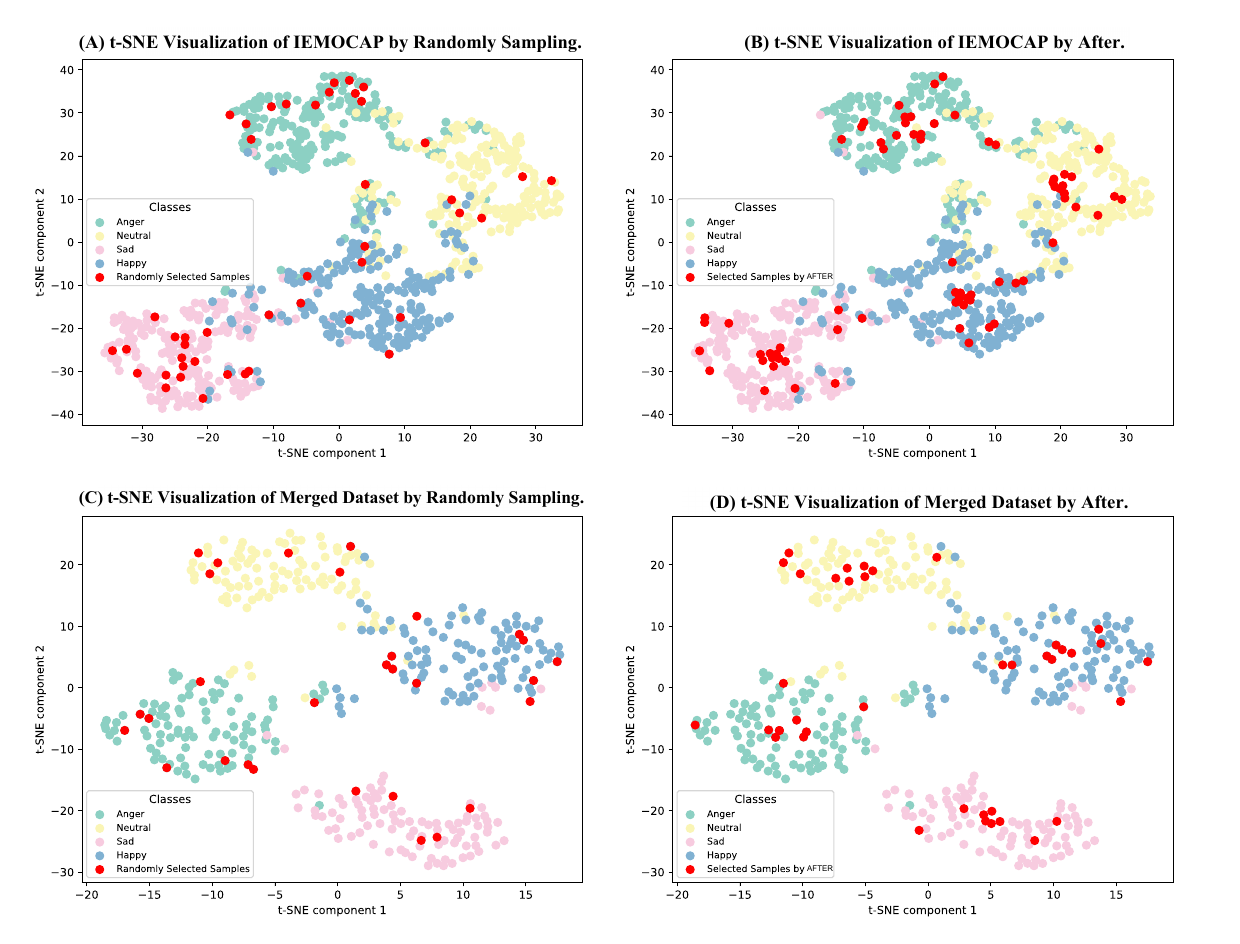}
\caption{t-SNE visualization of \textsc{After} and randomly sampled methods. The selected samples are represented with red colors on the IEMOCAP and Merged dataset by either randomly sampling or \textsc{After}.}
\label{visualization}
\end{figure*}

\subsection{Visualization of \textsc{After}}

As depicted in Figure~\ref{visualization}, we present qualitative comparisons of \textsc{After} with random sampling. 
Our observations indicate that \textsc{After} tends to select samples that are representative and uncertain.
Specifically, \textsc{After} selects samples near each clustering center, benefiting from the K-means initialization, which are the most representative samples of the entire datasets.
Using entropy as a criterion, \textsc{After} selects the most uncertain samples for labeling, which almost lie on the clustering boundaries. Based on previous experimental results, we discovered that only these selected representative and highly uncertain samples could achieve comparable or even superior performance compared to training with the entire datasets.
Additionally, the samples selected for training via random sampling are shown in Figure~\ref{visualization} (A) and (C). We found that random sampling tends to select outliers and redundant samples (some points overlap due to repeated selection).

\begin{figure*}[t]
\centering
\includegraphics[width=0.8\textwidth]{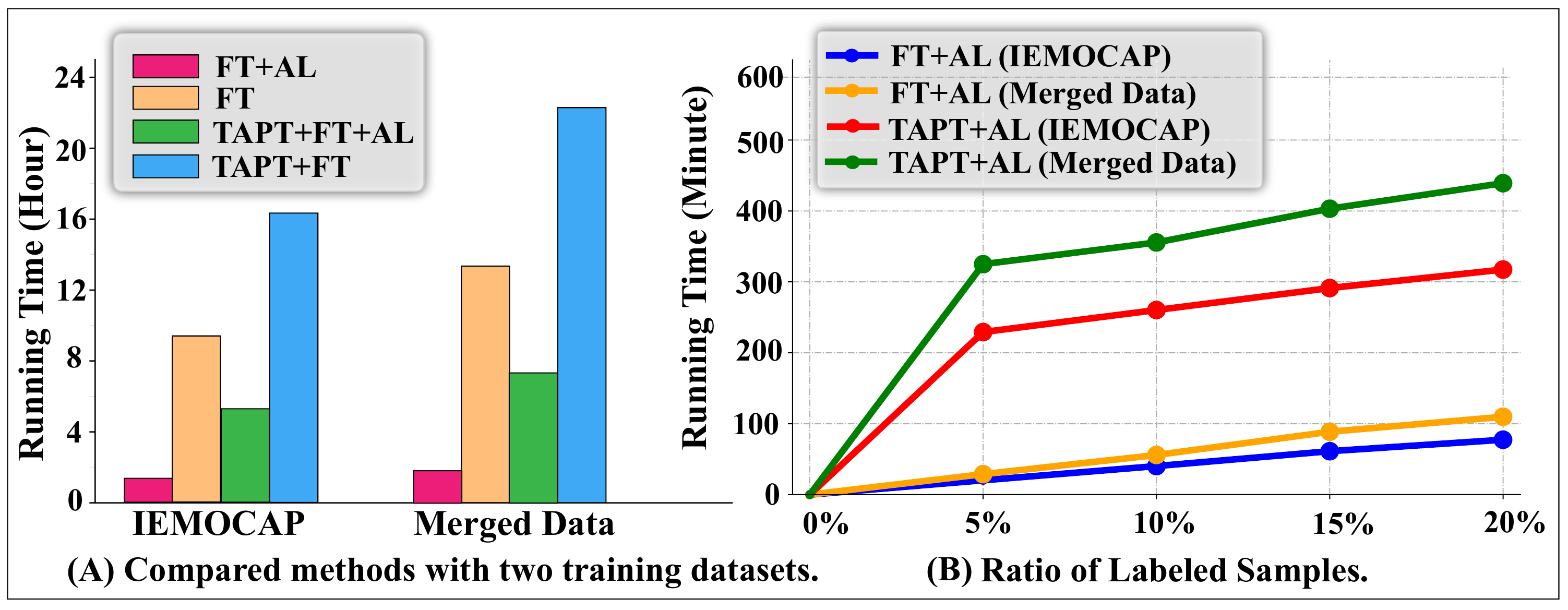}
\caption{ (A) Time Consumption Comparison and (B)  Relationship between ratio of labeled samples and time consumption.}
\label{time}
\end{figure*}

\begin{figure*}[h]
\centering
\includegraphics[width=0.8\textwidth]{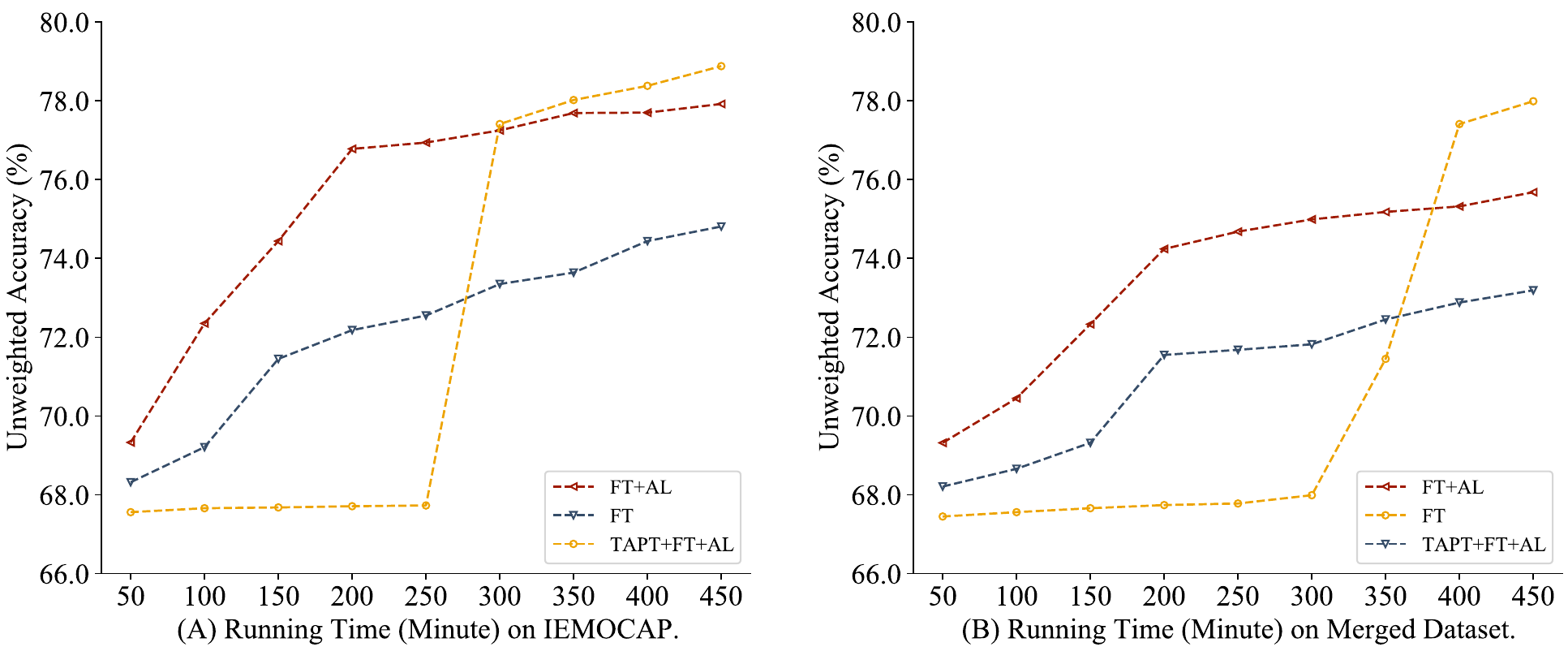}
\caption{A plot illustrating the efficiency of \textsc{After}. The x-axis represents the running time, while the y-axis indicates the unweighted classification accuracy.}
\label{time2}
\end{figure*}

\subsection{Time Consumption Comparison}
\label{running_time}

Figure~\ref{time} (A) demonstrates that FT+AL with 20\% labeled samples significantly reduced the time consumption of FT (fine-tuning on all labeled samples). Compared to TAPT+FT, TAPT+FT+AL significantly decreased the time consumption with the main cost incurred by TAPT. Additionally, the relationship between time consumption and the ratio of labeled samples is shown in Figure~\ref{time} (B). AL-based fine-tuning exhibited a linear increase in time consumption with sample size from 1\%$\sim$20\% (exponential growth from 30\%$\sim$100\% in Table~\ref{tab:tablenew}), indicating the efficiency of \textsc{After} and its potential to be applied in large-scale unlabeled real-world scenarios.

Figure~\ref{time2} illustrates the relationship between running time and unweighted accuracy of classification on both the IEMOCAP and Merged dataset.
We observe that FT+AL outperformed FT on both datasets, demonstrating the effectiveness of active learning.
Additionally, we also observe that TAPT+FT+AL underperformed FT and FT+AL within the time intervals of 0 to 250 minutes on IEMOCAP and 0 to 300 minutes on the Merged dataset, respectively. 
This is because TAPT requires time to adaptively pre-train the model using downstream unlabeled training datasets.
Following the TAPT process, TAPT+FT+AL (\textsc{After}) significantly outperformed the other two baselines on both datasets, thus demonstrating the effectiveness of our proposed method.

\begin{table}[h]
\centering
\caption{Unweighted and weighted accuracy on the Merged dataset. Entropy is used as an AL strategy for both HuBERT and wav2vec 2.0 backbones.}
\setlength{\tabcolsep}{4.3mm}{
\begin{tabular}{@{}lcccccccc@{}}
\toprule
\multicolumn{1}{l}{\multirow{3}{*}{\textbf{\textit{Methods}}}} 
& \multicolumn{4}{c}{\textbf{\small{HuBERT as backbone}}} 
& \multicolumn{4}{c}{\textbf{\small{wav2vec 2.0 as backbone}}} \\   
% & \multicolumn{2}{c}{\textbf{\small{Dublin}}}    
% & \multicolumn{2}{c}{\textbf{\small{arXiv}}}  \\ 
\cmidrule(lr){2-5} \cmidrule{6-9} 
% \cmidrule(lr){6-7} \cmidrule(lr){8-9}   
\multicolumn{1}{c}{}       & \multicolumn{2}{c}{\textbf{\small{FT+Random}}}                       & \multicolumn{2}{c}{\textbf{\small{\textsc{After}}}}  
& \multicolumn{2}{c}{\textbf{\small{FT+Random}}} 
& \multicolumn{2}{c}{\textbf{\small{\textsc{After}}}}                   \\
\cmidrule(lr){2-3} \cmidrule(lr){4-5}  \cmidrule(lr){6-7} \cmidrule{8-9} 
\multicolumn{1}{c}{} & UA $\uparrow$ & WA $\uparrow$ & UA $\uparrow$ & WA $\uparrow$  & UA $\uparrow$ & WA $\uparrow$  & UA $\uparrow$ & WA $\uparrow$ \\
\midrule
\midrule
{10\%}  & 52.28 & 50.58 & 71.22 & 68.83 & 50.82 & 48.96 & 71.45 & 69.01 \\
{20\%}           & 54.32 & 52.43
& 76.32   & 74.11  & 51.37 & 49.92 & 77.41   & 74.32 \\
{30\%}          &58.44 & 56.26      &77.54 & 75.12 & 52.37 & 50.18 & 78.20 & 75.16                \\
{40\%}          & 60.55 & 59.42           &78.23 &75.18 & 55.68 & 52.21 & 78.64 & 75.48               \\
{60\%}          & 63.38 & 61.43          &78.98 & 75.95 & 60.39 & 59.32 & \textbf{79.32} & \textbf{76.03}              \\
{80\%}          & 64.56 & 62.34          &79.24 & 75.81 & 58.34 & 56.72 & 79.26 & 75.92               \\
{100\%}                              & 63.45          & 61.28     & 78.89  & 79.35 & 57.21          & 54.12 & 79.15 & 75.94                    \\
\bottomrule
\end{tabular}}\label{ablation_studys_new}
\end{table}

\subsection{Adapting \textsc{After} with Different Pre-trained ASR Models}
\label{ASR-models}

Many SSL models have been proposed recently~\footnote{\url{https://www.isca-archive.org/interspeech_2023/index.html}}, such as  LABERT~\cite{fatehi23_interspeech} and DPHuBERT~\cite{peng23c_interspeech}.
To demonstrate the versatility of our proposed method, \textsc{After}, across various SSL models, we replaced wav2vec 2.0 with another widely used SSL model, HuBERT~\cite{DBLP:journals/taslp/HsuBTLSM21,DBLP:journals/corr/abs-2305-12442}, in conducting our experiments.

We first briefly introduce the HuBERT.  
HuBERT, also known as Hidden-Unit BERT, is a variant of BERT (Bidirectional Encoder Representations from Transformers) designed specifically for speech processing tasks. It leverages the powerful pre-training strategies of BERT while adapting its architecture to handle speech data efficiently.
Its core techniques are summarized as follows:

\begin{itemize}
\item \textbf{Architecture Adaptation}: HuBERT adapts the BERT architecture to process speech data. It modifies the input layer to accommodate raw audio waveforms and adjusts subsequent layers to handle sequential data inherent in speech signals. HuBERT incorporates transformer layers to capture contextual information from the extracted features. These layers enable the model to understand the temporal dependencies and nuances present in speech sequences.

\item \textbf{Feature Extraction}: Similar to traditional speech processing pipelines, HuBERT first extracts high-level features from raw audio using techniques like Mel-frequency cepstral coefficients or filter banks. These features capture important acoustic characteristics of the speech signal.

\item \textbf{Training Strategy}: HuBERT is pre-trained on large-scale unlabeled speech datasets using self-supervised learning techniques. During pre-training, the model learns to predict masked portions of the input sequence or to reconstruct corrupted segments, leveraging the bidirectional context provided by the transformer architecture. 
\end{itemize}

The TAPT process for HuBERT-based \textsc{After} follows a similar approach to that of  word2vec 2.0-based \textsc{After}. We first pre-train HuBERT using downstream speech emotion training datasets without using emotion labels, in an SSL manner. We randomly mask 20\% of tokens and reconstruct them using emotion recognition training datasets. Then, similar to theword2vec 2.0-based \textsc{After} method, we use K-means to select 1\% of samples for initialization. Finally, we iteratively query samples to train the HuBERT-based \textsc{After} model.

Experimental results are presented in Table~\ref{ablation_studys_new}. 
We have two observations:  (1) HuBERT-based fine-tuning with random sampling surpassed wav2vec 2.0-based fine-tuning with random sampling on the Merged datasets by nearly 2\%, showcasing HuBERT's effectiveness. This improvement can be attributed to HuBERT's tailored architecture for processing speech data, which may better capture subtle emotional cues. Additionally, HuBERT's pre-training objectives are more aligned with the demands of emotion recognition tasks.
(2) wav2vec 2.0-based \textsc{After} outperformed HuBERT-based \textsc{After}.
This is because wav2vec 2.0-based \textsc{After} pre-trains the model with contrastive loss and reconstructing loss, which better assists the SSL model in understanding the context of downstream datasets. 
These observations highlight the advantages and differences between HuBERT-based fine-tuning and wav2vec 2.0-based fine-tuning, as well as the impact of pre-training strategies on downstream performance.

%%%%%

\subsection{Extension of \textsc{After} to Multiple Annotators}
\label{multiple-annotators}

In this study, following previous works~\cite{DBLP:conf/icimcs/XuSZ13,zhang-etal-2022-survey}, we assumed that each sample is annotated with its ground truth labels. Thus, we first masked the labels on all training datasets and considered them as unlabeled data. Then, we unmasked the labels of some samples for training if these samples were selected by active learning.
However, in the real world, annotators with different knowledge, ages, genders, intuitions, backgrounds, and cultures~\cite{DBLP:conf/acllaw/BhardwajPSI10,dang2010comparison} may annotate the same sample differently.

Following previous studies, such as learning from the soft label~\cite{DBLP:conf/iccv/PetersonBGR19,DBLP:conf/hcomp/UmaFHPPP20,fornaciari-etal-2021-beyond} and learning from the hard label of individual annotators~\cite{cohn-specia-2013-modelling,DBLP:conf/aaai/RodriguesP18,10.1162/tacl_a_00449}, we extended our proposed method, \textsc{After}, to address the aforementioned real-world situation by suggesting the following potential solutions:
\begin{itemize}
\item (1) Individual-level Entropy (indi): We can measure the reliability of each annotator by calculating the individual-level entropy for each annotator. 
Given the prediction label for sample $x$ as $\textbf{z}^{a}=[z_{1}^{a},\cdots,z_{n}^{a}]$ by annotator $a$, the entropy can be calculated by 
\begin{equation}
    H_{indi}(p^{a}|x) = -\sum_{i=1}^{n} p^{a}_{i}(x)\log(p_{i}^{a}(x)),
\end{equation}
where $p_{i}^{a}(x) = \text{softmax}({z_{i}^{a}(x)})$. We select the ($instance, annotator$) pair with the highest entropy using: 
\begin{equation}
    argmax_{x \in U, a \in A} H_{indi}(p^{a}|x),
\end{equation}
where $U$ denotes the unlabeled set and $A$ denotes the annotator pool. 
%We compute entropy only for the remaining annotators who have not provided annotations for the instance.
\item (2) Group-level Entropy (group): Instead of focusing solely on individual uncertainty, we can query the data by considering the group-level uncertainty. To represent the uncertainty of the group on a sample, we calculate the entropy baseline based on the aggregation of each annotator's specific output. Therefore, we normalize and sum the logits of each annotator at the group level: $\mathbf{z}_{gropu} = [z_{1},\cdots z_{n}] =\sum_{a=1}^{|A|} \mathbf{z}_{norm}^{a},$ and calculate the group-level entropy as follows:
\begin{equation}
    H_{group}(x) = -\sum_{i=1}^{n} p_{i}(x) \log(p_{i}(x)),
\end{equation}
where $p_{i}(x) = \text{softmax} (z_{i}(x))$ and $|A|$ represent the number of annotators. We then query the data with the highest group-level uncertainty.
\item (3) Vote Variance (vote): Another method to measure the uncertainty among a group is by computing the variance of the votes. Given the prediction $y^{a}$  of annotator $a$, we calculate the vote variance as follows:
\begin{equation}
    \text{Var}(x) = \frac{1}{|A|} \sum_{a=1}^{|A|} (y^{a} -\mu)^{2},
\end{equation}
where $\mu = \frac{1}{|A|} \sum_{a=1}^{|A|} y^{a}$ and $|A|$ represents the number of annotators.
\item (4) Mixture of Group and individual Entropy (mix):  We also consider a variant that combines the group-level and individual-level entropy by simply adding the two $H_{mix} = H_{indi} + H_{group}$.
\end{itemize}

\subsection{Extension of \textsc{After} with Soft-labels}
\label{soft-labels}

As the number of emotions increases, the difference in the results depending on the annotator becomes more pronounced. Generally, soft labels are used to address this issue. 
We conducted additional experiments on the IEMOCAP dataset, where each utterance was labeled by three human annotators. Furthermore, each annotator was allowed to choose more than one categorical label if they felt it necessary~\cite{BussoBLKMKCLN08}.

To simulate three different annotators, following \cite{DBLP:conf/ijcnn/FayekLC16}, we trained three separate DNNs on the IEMOCAP dataset using the hard labels from each annotator. Specifically, each DNN architecture contained seven feed-forward fully-connected layers and adopted ReLU as the activation function. The input layer's dimensionality was 2,624 (64 frames $\times$ 41 coefficients per frame) and the output layer is a four-way softmax layer, which produced the posterior class probabilities. We used the cross-entropy loss for the emotion recognition of each classifier:
\begin{equation}
    \mathcal{L}_{ce} = -\frac{1}{N} \sum_{i=1}^{N}\sum_{j=1}^{c} y^{j}_{i} \log{ (\hat{y}_{i}^{j})}, \label{cm3}
\end{equation}
where $c$ is the number of emotion classes,  $N$ denotes the total number of samples, $\hat{y}_{i}^{j}$ stands for the $i$-th predicted label, and $y_{i}^{j}$ represents the $i$-th ground-truth label for the $j$-th class. For more detailed implementations, please refer to \cite{DBLP:conf/ijcnn/FayekLC16}.

Then we obtained the soft labels for each speech sample by:
\begin{equation}
    s = \frac{\sum_{a=1}^{A}h^{(n)}}{\sum_{i=1}^{c}\sum_{a=1}^{A}h^{(n)}},
\end{equation}
where $s$ is a $c$-dimensional vector of soft labels, $h^{(n)}$ is a $c$-dimensional vector of one-of-$c$ hard labels encoded from the $n$-$th$ annotator, and A is the number of annotators. Table~\ref{example-soft-labels} illustrates several annotation examples from the IEMOCAP database labeled by three annotators and their corresponding labels to alleviate ambiguity.

%%%%%%%%%%%%%%% 
% How to define this issue is a problem.}\\ 

% \begin{equation}\label{layer-1}
%     \hat{\mathbf{y}}^{(L)} = \frac{\exp (\mathbf{z}^{(L)})}{\sum_{1}^{c} \exp(\mathbf{z}^{(L)})}
% \end{equation}
% \begin{equation}\label{layer-2}
%     z^{(L)} = y^{(L-1)}W^{(L)} + b^{(L)},
% \end{equation}
% where $y^{(L)}$ is a vector of normalized class probabilities, $z^{(L)}$ is the input to layer L computed as in Eq.(\ref{layer-2}), $y^{(L-1)}$ is the output of the final hidden layer, $L=7$ is the total number of layers, $W^{(L)}$ and $b^{{(L)}}$ are a matrix of weights and a vector of biases respectively, and $c=4$ is the number of output classes.

\begin{table}[h]
\centering
\caption{Examples of soft labels for IEMOCAP with three annotators. Annotation are in the form of (Annotator 1, Annotator 2, Annotator 3). Hard/Soft labels are in the form of [Anger, Happiness, Neutral, Sadness].}
\setlength{\tabcolsep}{16mm}{
\begin{tabular}{@{}lcc@{}}
\toprule
{Annotation}                         
& Hard Label	         & Soft Label                    \\
\midrule
\midrule
(Anger, Anger, Anger)               & [1,0,0,0]          & [1,0,0,0]                            \\
(Happiness, Neutral, Neutral)   & [0,0,1,0]          & [0,0.33,0.66,0]      \\
(Sadness, Sadness, Sadness;Neutral)                   & [0,0,0,1]          & [0,0,0.25,0.75]                                    \\
\bottomrule
\end{tabular}
}\label{example-soft-labels}
\end{table}

After obtaining the soft labels from the IEMOCAP datasets, \textsc{After} also measures the uncertainty of each sample $\mathbf{x}_{i}$ as follows:
\begin{equation} \textrm{Entropy}(\mathbf{x}_{i}) = -\sum_{j=1}^{c}P(\hat{y}_{j}|\mathbf{x}_{i})\textrm{log}P(\hat{y}_{j}|\mathbf{x}_{i}), \end{equation} 
where $c$ is the number of emotional classes and $P(\hat{y}_{j}|\mathbf{x}_{i})$ represents the predicted probability of $\mathbf{x}_{i}$ for the $j$-th emotion. Following \cite{DBLP:conf/ijcnn/FayekLC16}, the classifier outputs the class with the highest posterior probability during evaluation. Experimental results in Table~\ref{performance-soft-label} demonstrate that \textsc{After} outperformed all baselines even with soft labels, indicating its capability to handle real-world scenarios with complex soft-labeled emotions.

\begin{table}[h]
\centering
\caption{Overall performance comparison on four emotion categories. 
\textsc{After} adopted Entropy+Clustering and selected 20\% samples for fine-tuning. Baselines use all samples from each corresponding dataset for training. THe symbol $\dag$ indicates that \textsc{After} significantly surpassed all baselines with $p<0.05$ according to the t-test.}
\setlength{\tabcolsep}{8.5mm}{
\begin{tabular}{@{}lcccccc@{}}
\toprule
\multicolumn{1}{l}{\multirow{2}{*}{\textbf{\textit{Methods}}}} 
& \multicolumn{2}{c}{\textbf{\small{IEMOCAP (hard-label)}}} 
& \multicolumn{2}{c}{\textbf{\small{IEMOCAP (soft-label)}}} 
\\   
% & \multicolumn{2}{c}{\textbf{\small{Dublin}}}    
% & \multicolumn{2}{c}{\textbf{\small{arXiv}}}  \\ 
\cmidrule(lr){2-3} \cmidrule(lr){4-5} \cmidrule(lr){6-7} 
% \cmidrule(lr){6-7} \cmidrule(lr){8-9}   
\multicolumn{1}{c}{}                         & UA $\uparrow$	         & WA  $\uparrow$       & UA  $\uparrow$        & WA $\uparrow$                 \\
\midrule
\midrule
GLAM [2022]                   & 74.01          & 72.98          & 68.15          & 64.33                    \\
LSSED [2021]                    & 73.09          & 68.35          & 62.23          & 61.38                   \\
RH-emo [2022]                    & 68.26          & 67.35          & 61.35          & 59.17                    \\
Pseudo-TAPT [2022]                    & 74.30          & 70.26          & 69.98          & 68.23   \\
w2v2-L-r-12 [2023] &74.28 &70.23 &70.31 &69.24 \\
\midrule
\textbf{\textsc{After}}                                      & $ \,\,\textbf{76.07}^{\dag}$ & $ \,\,\textbf{73.24}^{\dag}$  & $ \,\,\textbf{73.37}^{\dag}$ &$ \,\,\textbf{72.96}^{\dag}$  \\
\bottomrule
\end{tabular}
}\label{performance-soft-label}
\end{table}

\section{Limitations}\label{sec:limitation}

Although \textsc{After} achieves SOTA performances on IEMOCAP and the Merged dataset, there are still some limitations in this study.
(1) Its performance on larger-scale and more heterogeneous real-world data remains unclear.
(2) Another limitation of \textsc{After} is the time-consuming process of calculating the entropy of each sample in each AL iteration. Additionally, we have only explored five of the most commonly used AL strategies in this study, leaving better strategies unexplored.
(3) Annotation of emotions varies greatly depending on various factors. While this study assumes that annotators can always provide ground-truth labels, 
current experimental settings may not sufficiently simulate real-world human-in-the-loop situations involving multiple annotators.
Although we propose a potential solution for handling multiple annotators in real-world scenarios, we lack evaluation for it. We plan to address this issue by enlisting human annotators to label samples in future evaluations.
(3) We need to apply our methods to more complicated scenes, such as social networks with clustering technique~\cite{li2021identification,li2023temporal,li2021joint} and clinical patient emotion detection~\cite{8983045,9388900}. 
(4) Finally,  \textsc{After} is designed specifically for multimodal emotion recognition tasks, which are not as straightforward and generalizable as language models~\cite{Roberta,kenton2019bert} or image processing techniques~\cite{lecun1995convolutional}. 

\section{Comparison with the Previous Conference Version}\label{sec:previous}

In this section, we primarily delineate three key distinctions between this version and the conference version of our research. 
(1) We introduced novel methodologies by extending \textsc{After} to accommodate more complex real-world scenarios. Specifically, we first explored various initialization methods for \textsc{After} in Section~\ref{initialized-methods}. Subsequently, we investigated the impact of integrating \textsc{After} with different pre-trained ASR models in Section~\ref{ASR-models}. Finally, 
we extended \textsc{After} to encompass more complex scenarios involving multiple annotators in Section~\ref{multiple-annotators} and soft-labels in Section~\ref{soft-labels}.
(2) We redefined the motivation behind this study in Section~\ref{sec:Introduction}, as well as redefined the meaning of ``noisy'', ``heterogeneous'', and ``real-world''. 
To better simulate "real-world" scenarios, we created a new dataset named ``Merged-2 Dataset'', comprising both acted and spontaneous datasets. 
Additionally, we created another novel dataset, ``Merged-3 Dataset'', incorporating two acted speech datasets (EmoDB, RAVDESS), and two spontaneous datasets (AFEW5.0, BAUM-1s) across seven emotional categories.
(3) We conducted additional experiments and provided in-depth analysis.
Firstly, we conducted experiments to analyze the effects of initialization methods in Section~\ref{initialized-methods} and the influence of various pre-trained ASR models in Section~\ref{ASR-models}. 
Subsequently, we conducted experiments on the newly added datasets ``Merged-2 Dataset'' and ``Merged-3 Dataset'', accompanied by detailed analysis in Section~\ref{sec.compare with sota}. 
Finally, we added additional ablation studies in Section~\ref{ablation_study} and compared time consumption in Section~\ref{running_time}. Additionally, we explored the applicability of our methods to multiple annotators and soft-labels in Section~\ref{soft-labels}.

\section{Conclusion}\label{sec:conclusion}

In this study, we investigated unsupervised TAPT and the AL-based fine-tuning strategy to improve the performance of SER. 
To extend SER applications to real-world scenarios, we created three large-scale noisy and heterogeneous datasets, and we used TAPT to bridge the information gap between pre-trained and the target SER task. 
Experimental findings demonstrate that \textsc{After} significantly improved performance and reduced time consumption.
In our future work, we plan to create larger-scale speech emotion recognition datasets for testing in the speech domain. Furthermore, we aim to explore and design more effective and efficient active learning strategies tailored to the SER task, aiming to minimize time consumption.
Finally, we would like to propose a more general framework that extends beyond SER, focusing on a wider range of speech or language-related tasks.

\bibliography{e_yourrefs}
\bibliographystyle{unsrt}

\end{document}